\documentclass[aps,prb,superscriptaddress,twocolumn,floatfix]{revtex4-1}
\usepackage{graphicx}
\usepackage{amssymb}
\usepackage{amsfonts}
\usepackage{amsmath}
\usepackage{textcomp}

\begin{document}

\title{Type-I superconductivity in PdTe$_2$ probed by $\mu$SR}

\author{H. Leng} \email{h.leng@uva.nl}\affiliation{Van der Waals - Zeeman Institute, University of Amsterdam, Science Park 904, 1098 XH Amsterdam, The Netherlands}
\author{J.-C. Orain} \affiliation{Laboratory for Muon-Spin Spectroscopy, Paul Scherrer Institute, 5232 Villigen PSI, Switzerland}
\author{A. Amato} \affiliation{Laboratory for Muon-Spin Spectroscopy, Paul Scherrer Institute, 5232 Villigen PSI, Switzerland}
\author{Y. K. Huang} \affiliation{Van der Waals - Zeeman Institute, University of Amsterdam, Science Park 904, 1098 XH Amsterdam, The Netherlands}
\author{A. de Visser} \email{a.devisser@uva.nl} \affiliation{Van der Waals - Zeeman Institute, University of Amsterdam, Science Park 904, 1098 XH Amsterdam, The Netherlands}
\date{\today}

\begin{abstract}
The Dirac semimetal PdTe$_2$ was recently reported to be a type-I superconductor with $T_c = 1.64$~K and a critical field $\mu_0H_c = 13.6$~mT. Since type-I superconductivity is unexpected for binary compounds, we have conducted muon spin rotation experiments to probe the superconducting phase on the microscopic scale via its intermediate state. For crystals with a  finite demagnetization factor, $N$, the intermediate state forms in applied fields $(1-N)H_c < H_a < H_c$. We have carried out transverse field muon spin rotation measurements on a thin disk-like crystal with the field perpendicular to ($N_{\perp}=0.86$) and in the plane ($N_{\parallel}=0.08$) of the disk. By analysing the $\mu$SR signal we find that the volume fraction of the normal domains grows quasi-linearly with applied field at the expense of the Meissner domain fraction. This then provides solid evidence for the intermediate state and type-I superconductivity in the bulk of PdTe$_2$.

\end{abstract}

\maketitle
\section{Introduction}
The large family of layered transition metal dichalcogenides is extensively studied because of their fascinating electronic properties. One of the modern-day research interests is a non-trivial nature of the electronic band structure, which may result in topology driven quantum states. Density functional calculations show, for instance, that selected transition metal dichalcogenides host generic  three-dimensional type-II Dirac fermion states~\cite{Soluyanov2015,Huang2016,Yan2017,Bahramy2018}. In a type-II Dirac semimetal the Dirac cone, which embodies the linear energy dispersion, is tilted, and the Hamiltonian breaks Lorentz invariance~\cite{Soluyanov2015}. Here we focus on the exemplary material PdTe$_2$. Extensive electronic structure calculations combined with angle resolved photoemmission spectroscopy (ARPES) demonstrate a type-II Dirac semimetallic state with the Dirac point at $\sim 0.6$~eV below the Fermi energy~\cite{Liu2015a,Fei2017,Noh2017,Bahramy2018,Clark2017}. Another interesting property of PdTe$_2$ is that it superconducts below $T_c = 1.6$~K~\cite{Guggenheim1961}. In a type-II Dirac semimetal the Dirac point is the touching point of the electron and hole pockets and a nearly flat band may form near the Fermi level. This could promote superconductivity, which in turn prompts the question whether superconductivity has a topological nature~\cite{Fei2017,Leng2017}.

In a recent paper Leng \textit{et al.}~\cite{Leng2017} reported a magnetic and transport study on single crystalline PdTe$_2$ and concluded superconductivity shows type-I behavior. This result is surprising, because binary compounds when superconducting exhibit in general type-II behavior. Until today this rare phenomenon has been documented convincingly for about a dozen binary or ternary compounds only (see Ref.~\onlinecite{Peets2019}). In the case of PdTe$_2$ evidence for type-I behavior is provided by (\textit{i}) the dc-magnetization curves as function of the applied field, $M(H_a)$, that show the presence of the intermediate state between $(1-N)H_c < H_a < H_c$, where $N$ is the demagnetization factor and $H_c$ the critical field with $\mu_0 H_c(0) = 13.6$~mT, (\textit{ii}) the differential paramagnetic effect (DPE), that shows up as a peak in the ac-susceptibility in applied dc-field, just below $T_c$, and (\textit{iii}) the quadratic temperature variation of the thermodynamic critical field $H_c(T) = H_c(0)[1-(T/T_c)^2]$. The value of the Ginzburg-Landau parameter $\kappa = \lambda / \xi $, where $\lambda$ is the magnetic penetration depth and $\xi$ the superconducting coherence length, amounts to 0.09-0.29~\cite{Leng2017,Salis2018} and is smaller than $1/\sqrt{2}$, the boundary value for type-I and type-II behavior. The superconducting phase has further been characterized by heat capacity~\cite{Amit2018}, scannning tunneling microscopy/spectroscopy (STM/STS)~\cite{Das2018,Clark2017,Sirohi2019}, and magnetic penetration depth measurements~\cite{Teknowijoyo2018,Salis2018}. The specific heat data confirm conventional weak-coupling Bardeen-Cooper- Schrieffer superconductivity with a ratio $\Delta c/ \gamma T_c \approx 1.52$, which is close to the weak-coupling value 1.43. Here $\Delta c$ is the size of the step in the specific heat at $T_c$ and $\gamma$ the Sommerfeld coefficient. The STM/STS spectra taken in zero magnetic field point to a fully-gapped superconducting state, without any in-gap states. Finally, the magnetic penetration depth, $\lambda (T)$, shows an exponential temperature variation for $T/T_c < 0.4$ consistent with a fully-gapped superconducting state.

Nonetheless, several curious features have come to the fore in the superconducting state of PdTe$_2$. First of all, ac-susceptibility measurements in a small driving field have revealed large screening signals in applied dc-fields $H_a > H_c$ (Ref.~\onlinecite{Leng2017}) (here $H_a$ is directed along the $a$-axis). This has been attributed to superconductivity of the surface sheath~\cite{Leng2017}. Screening persists up to the critical field $\mu_0 H_c^S (T \rightarrow 0) = 34.9$~mT. Surface superconductivity is not of the standard Saint-James - de Gennes type, which has a critical field $H_{c3} = 2.39 \times \kappa H_c$ (Ref.~\onlinecite{Saint-James&deGennes1963}). In fact when $\kappa < 0.42$, $H_{c3} < H_c$ and Saint-James - de Gennes surface superconductivity should not occur. This opens up the possibility that superconductivity of the surface layer has a different nature and originates from the topological surface states that were detected by ARPES~\cite{Liu2015a,Noh2017}. Another striking feature is that electrical resistance measurements reveal superconductivity to survive up to fields that are much higher, typically $\mu_0 H_c ^R (0) =0.3$~T $\gg \mu_0 H_c^S(0) > \mu_0 H_c(0)$~(Ref.\onlinecite{Leng2017}). The resulting complex phase diagram in the $H-T$ plane shows some similarities with the diagrams reported for the superconductors LaRhSi$_{3}$~\cite{Kimura2016} and ZrB$_{12}$~\cite{Wang2005}. However, in these cases the unusual diagram is attributed to a field-induced change from type-I to type-II superconductivity below a conversion temperature $T^* < T_c$. These materials are called type-II/1 superconductors, and have a $\kappa$-value close to 1/$\sqrt{2}$~(Ref.~\onlinecite{Auer&Ullmaier1973}).

Another puzzling aspect comes from STM/STS measurements in applied dc fields. Das \textit{et al.}~\cite{Das2018} have investigated the closure of the gap for a field along the $c$-axis at $T/T_c = 0.23$ and find that the superconducting gap predominantly is suppressed at a critical field $\mu_0 H_c(0) \approx 25$~mT. However, they also find regions on the surface of the crystal where significantly larger fields are required to suppress superconductivity, typically in the range 1-4 T. These STM/STS results were taken a step further by Sirohi \textit{et al.}~\cite{Sirohi2019} who reported a distinct behavior in the spectra taken in the low and high $H_c$ regions. They concluded that the observed spatial distribution of critical fields is due to mixed type-I and type-II superconducting behavior, which in turn stems from electronic inhomogeneities visible in the spectra in the normal state. A third STM/STS characterization was carried out by Clark \textit{et al}.~\cite{Clark2017} Since these authors observe a vortex core in a field of 7~mT they claim PdTe$_2$ is a type-II superconductor, and report an upper field critical field $\mu_0 H_{c2} = 20$~mT. We remark, that in the STM/STS work reported so far, evidence of an Abrikosov vortex lattice has not been produced. More recently, mechanical and soft point contact spectroscopy (PCS) data were also taken as evidence for mixed type-I and type-I superconductivity on the surface~\cite{Le2019}. A possible issue in all these experiments is that the applied field was directed perpendicular to a flat crystal, which involves a large demagnetization factor and the formation of the field-induced intermediate state. This possibility has not been addressed in the aforementioned STM/STS papers.

These conflicting results warrant the investigation of the superconducting phase of PdTe$_2$ on the microscopic scale. For this the $\mu$SR technique is extremely well suited, because it is a local probe which permits to determine whether regions with distinct magnetic properties are present in the crystal~\cite{Amato1997,Yaounc&Dalmas2011}. $\mu$SR is also a well-established technique to measure the penetration depth of type-II superconductors~\cite{Blundell1999}. In the transverse field configuration the precession of the muon ($\mu ^+$) spin is damped by the local field distribution of the vortex lattice. From the resulting Gaussian damping rate, $\sigma(T)$,  the magnetic penetration depth, $\lambda(T)$, can be derived. In a type-I superconductor in the Meissner phase, the application of a transverse field will not give rise to precession of the $\mu ^+$ spin because the magnetic induction in the crystal is zero. However, for applied fields larger than $(1-N)H_c$ the intermediate state is generated and a macroscopic phase separation occurs in Meissner and normal state domains. The field in the normal regions is equal to the critical field $H_c$. Consequently, $\mu ^+$ spin precession will occur in the normal-phase fraction of the crystal. By fitting the $\mu$SR signal with the appropriate muon depolarization function, one can determine the Meissner and normal phase fractions in the crystal.

Although a powerful technique, $\mu$SR on type-I superconductors has not been explored in much detail. Studies of the intermediate state in elemental superconductors are scarce and concise~\cite{Gladisch1979,Grebinnik1980,Egerov2001,Aegerter2003,Kozhevnikov2018,Khasanov2019,Karl2019}. The most recent work by Karl \textit{et al.} (Ref.~\onlinecite{Karl2019}), however, presents a comprehensive review of the technique and an in-depth analysis of the $\mu$SR signal in the intermediate phase of a $\beta$-Sn sample. Binary and ternary compounds that have been scrutinized for type-I superconductivity include LaNiSn~\cite{Drew2006}, LaRhSi$_3$~\cite{Anand2011}, LaIrSi$_3$~\cite{Anand2014}, LaPdSi$_3$~\cite{Smidman2014}, and very recently AuBe~\cite{Singh2019,Beare2019}.

Here we report transverse field muon spin rotation measurements in the superconducting phase of PdTe$_2$. Experiments were performed on a thin disk-like crystal in two configurations: (\textit{i}) with the field perpendicular to the plane of the disk ($N_{\perp}=0.86$) and (\textit{ii}) with the field in the plane of the disk ($N_{\parallel}=0.08$). By analysing the asymmetry of the $\mu$SR signal we find that the normal phase volume fraction grows quasi-linearly with applied field at the expense of the Meissner phase fraction. This provides solid evidence for the intermediate state and type-I superconductivity in the bulk of our PdTe$_2$ crystal.

\section{Experiment}

The PdTe$_2$ crystal used for the $\mu$SR experiment was taken from a single-crystalline boule prepared by the modified Bridgman technique\cite{Lyons1976}. Its single-crystalline nature was checked by Laue backscattering. Powder X-ray diffraction confirmed the trigonal CdI$_2$ structure (spacegroup $P\bar{3}m1$). Scanning electron microscopy (SEM) with energy dispersive X-ray (EDX) spectroscopy showed the proper 1:2 stoichiometry within the experimental resolution of 0.5\%. The superconducting properties of small crystals cut from the single-crystalline boule were measured by dc-magnetization and ac-susceptibility~\cite{Leng2017}. The Meissner volume fraction for a bar-shaped crystal cut along the $a$-axis, and $H_a \parallel a$, amounts to 93\% after correcting for demagnetization effects~\cite{Leng2017}. The crystal used in the present experiment is cut from the same region of the single-crystalline boule and has a disk-like shape, with the $c$-axis perpendicular to the plane of the disk. Its thickness equals 0.65~mm and the diameter is 10.0~mm. However, a small piece was removed and cut from the disk along the $a$-axis, which reduced the size in the perpendicular $a^*$-direction ($\perp a$) to 6.8~mm. This causes additional field inhomogeneities near the edges of the sample, notably for the configuration with the field in the plane of the disk. It also thwarts a precise calculation of the demagnetization factors. With appropriate approximations of the sample shape the estimated values are $N_{\perp} =0.86\pm0.02$ and $N_{\parallel}= 0.08\pm0.03$~\cite{Chen1991,Pardo2004}. These values have been calculated for a completely diamagnetic state, $\chi = -1$.

The crystal was attached with its flat surface utilizing vacuum grease (Apiezon N) to a thin copper foil that is supported by a fork-shaped copper holder. A thin layer of Kapton foil was wrapped around the sample and holder to mechanically fix the crystal. The holder was attached to the cold finger of a helium-3 refrigerator (HELIOX, Oxford Instruments) and $\mu$SR spectra were taken in the temperature range $T=0.25 - 5$~K. The crystal is oriented with its large surface perpendicular to the muon beam and the area for the implanted muons is $\sim$55~mm$^2$.

Muon spin rotation ($\mu$SR) experiments were carried out with the Multi Purpose Surface Muon Instrument DOLLY installed at the $\pi $E1 beamline at the S$\mu$S facility of the Paul Scherrer Institute. The technique employs the decay probability of  spin-polarized muons that are implanted in the crystal. In the case of PdTe$_2$ (density 8.3~g/cm$^3$) the muons typically penetrate over a distance of 133$\pm26~\mu$m, and thus probe the bulk of the crystal. In the presence of a local or applied field at the muon stopping site the muon spin will precess around the field direction with an angular frequency $\omega _{\mu} = \gamma_{\mu} B_{loc}$, where $\gamma_{\mu}$ is the muon gyromagnetic ratio ($\gamma_{\mu} /2 \pi = 135.5~$MHz/T). The subsequent asymmetric decay process is monitored by counting the emitted positrons by scintillation detectors that are placed at opposite directions in the muon-spin precession plane~\cite{Amato1997,Yaounc&Dalmas2011,Blundell1999}. The parameter of interest is the muon spin asymmetry function, $A(t)$, which is determined by calculating $A(t) =(N_1 (t) - \alpha N_2 (t))/(N_1 (t) + \alpha N_2 (t) )$, where $N_1 (t)$ and $N_2 (t)$ are the positron counts of the two opposite detectors, and $\alpha$ is a calibration constant. In our case $\alpha$ is close to 1.

Transverse field (TF) experiments were performed with the magnetic field applied parallel and perpendicular to the crystal plane. In the first configuration the muon spin is along the beam direction, the field in the horizontal plane at right angles to the beam (and in the plane of the disk, $N= N_{\parallel}$), and the decay positrons are detected in the backward and forward counters. In the second case the beam-line is operated in the muon spin-rotated mode, the applied field is along the beam direction (perpendicular to the plane of the disk, $N= N_{\perp}$), and the decay positrons are collected in the left and right counters. In the spin-rotated mode the muon spin is directed $\sim 45^\circ$ out of the horizontal plane. This results is a reduced asymmetry function ($A \approx 0.18$) with respect to the full asymmetry ($A \approx 0.23$) in the non-spin-rotated mode.  The $\mu$SR time spectra were analysed with the software packages WIMDA~\cite{Pratt2000} and MUSRFIT~\cite{Suter&Wojek2012}.

\begin{figure}
  \centering
  \includegraphics[width=8cm]{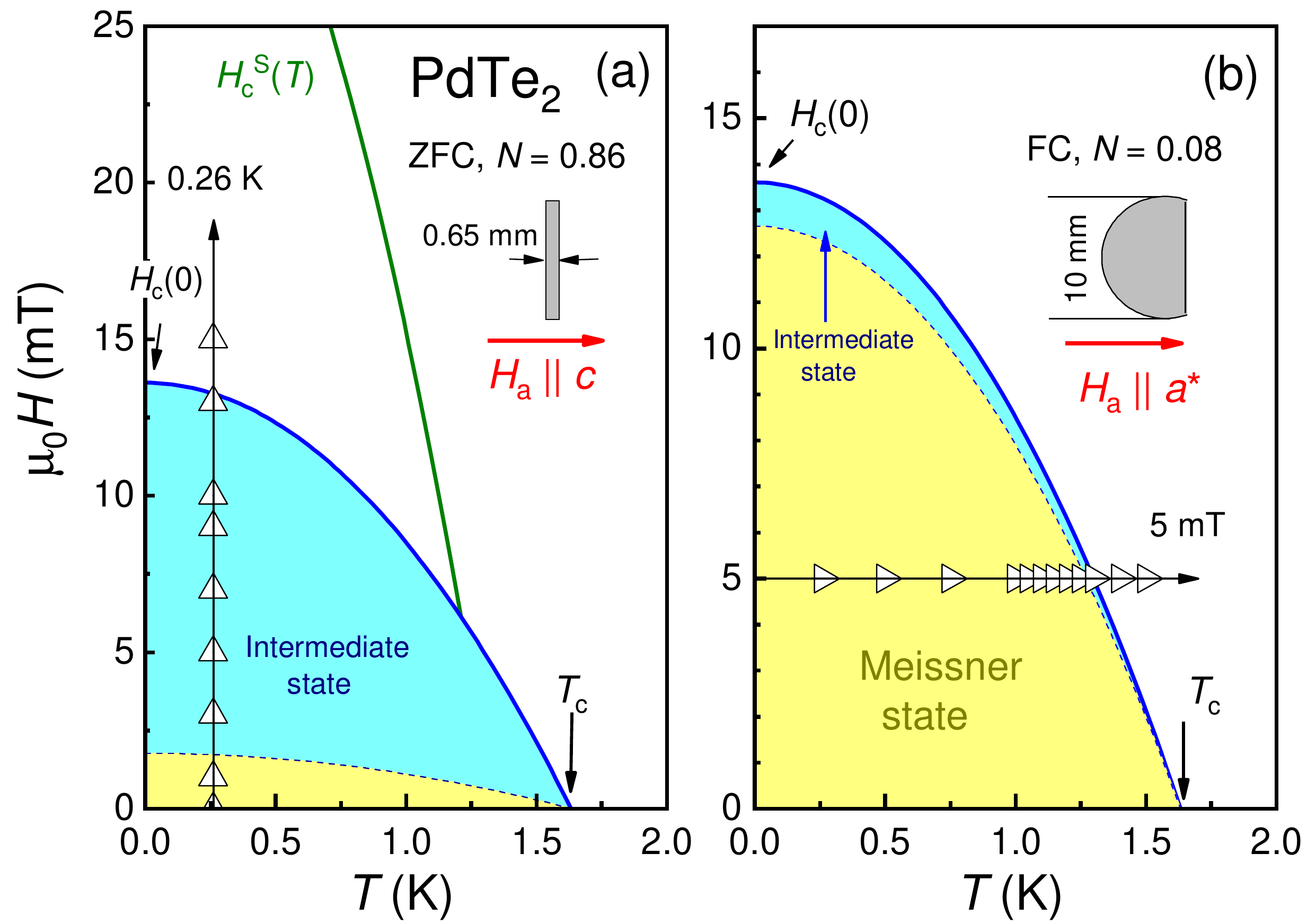}
  \caption{Field and temperature scan procedure of the superconducting phase diagram of PdTe$_2$ reported in Ref.~\onlinecite{Leng2017}. The blue colored area indicates the intermediate phase, and the yellow area the Meissner phase. (a) After zero field cooling (ZFC) down to $T=0.26$~K, spectra were recorded by increasing the field $H_a \parallel c$ step-wise at values denoted by the up-triangles. (b) After cooling down to 0.26~K in a field $H_a \parallel a^*$ of 5~mT, spectra were recorded at the temperatures indicated by the side triangles. In the upper part of (a) and (b) the sample and field geometry are sketched. The solid green line in (a) indicates the region below which surface superconductivity is observed~\cite{Leng2017}. Note the vertical scale is different in (a) and (b).}
  \label{procedure}
\end{figure}

\section{Results and Analysis}

In order to investigate the presence of the intermediate state we have scanned the superconducting phase diagram as depicted in Fig.~\ref{procedure}. In Fig.~\ref{procedure}(a) we show the case where the sample is slowly cooled in zero field (ZFC) after which the field, directed perpendicular to the plane of the disk, is increased in eight steps to a value $H_a > H_c$. In this case the intermediate state covers a large region of the phase diagram. In Fig.\ref{procedure}(b) we show the case where the sample is cooled in 5~mT (FC), applied in the plane of  the disk, after which the temperature is raised in eleven steps to $T > T_c $ (at 5~mT). In this case the intermediate state  region is expected to be small.

\subsection{Field perpendicular to the plane of the disk}

\begin{figure}
  \centering
  \includegraphics[width=7cm]{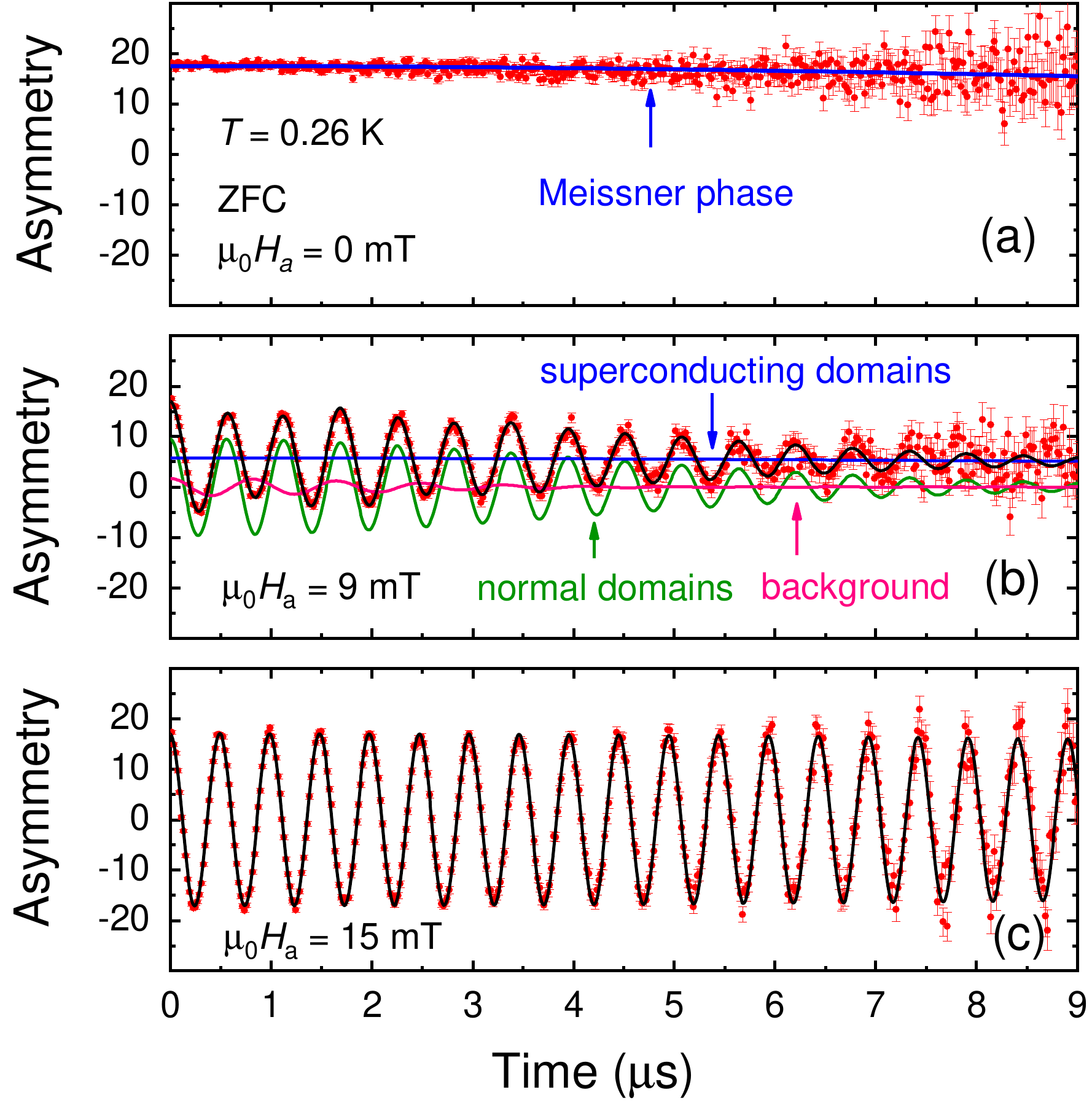}
  \caption{$\mu$SR spectra collected at $T= 0.26$~K in ZF and in applied fields of 9~mT and 15~mT directed perpendicular to the sample plane. (a) Zero-field. The solid blue line is a fit to the Guassian Kubo-Toyabe function Eq.~\ref{Gaussian Kubo-Toyabe function}. (b) TF = 9~mT.  The black line is a fit to the three component function Eq.~\ref{total fitting function}. The different components, due to superconducting domains, normal domains and background, are shown by the solid blue, green and pink lines, respectively. (c) TF = 15~mT. The black solid line is a fit to the depolarization function Eq.~\ref{normal}. See text for fit details.  }
  \label{spectra_260mK}
\end{figure}

In Fig.~\ref{spectra_260mK} we show three typical TF $\mu$SR spectra at $T=0.26$~K recorded during step-wise increasing the field to 15~mT. In panel (a) no field is applied and muon spin precession is absent, the muons probe the Meissner phase. In panel (b) the applied field is raised to 9~mT. Now a clear spin precession is visible, but with a reduced asymmetry. The superconducting volume has shrunk. The spin precession frequency corresponds to a local field  $B_{loc} = 13.0$~mT, which is equal to $\mu_0 H_c$ at 0.26~K. This shows the sample is in the intermediate state. Lastly, in panel (c) the field is raised to 15~mT $ > \mu_0 H_c$ and all muons show a precession frequency corresponding to $B_a = B_{loc} = 15$~mT, as expected in the normal state.

The $\mu$SR response $A(t)=AP(t)$, where $P(t)$ is the muon depolarization function, in panel (a) of Fig.~\ref{spectra_260mK} is well described by a Gaussian Kubo-Toyabe function
\begin{equation}\label{Gaussian Kubo-Toyabe function}
A_{KG}(t)= A_{0}[\frac{1}{3}+\frac{2}{3}(1-\sigma_{KG}^{2}t^{2} \exp (-\frac{1}{2} \sigma_{KG}^2 t^2))]
\end{equation}

\noindent
Here A$_0$ is the initial asymmetry and $\sigma _{KG}$ the depolarization rate. The fit is shown in panel (a) by the solid blue line. The fit parameters are $A_0 = 17.6$ and $\sigma _{KG} = 0.05~\mu$s$^{-1}$. The small depolarization rate is attributed to a Gaussian distribution of static nuclear moments. In the normal phase, panel (c), the $\mu$SR response is best fitted with the function (solid black line):
\begin{equation}\label{normal}
A_N(t)=A_{0}\exp (-\frac{1}{2} \sigma_{N}^2 t^2)\cos (\gamma_{\mu}B_a t+\phi_{N})
\end{equation}

\noindent
where $\sigma_{N}$ is a Gaussian damping rate, $B_a$ the applied field and $\phi_N$ a phase factor. The fit parameters are $A_0$= 17.4 and $\sigma _{N} = 0.04~\mu$s$^{-1}$. The small damping rate is attributed to the field distribution of nuclear moments as well, which is considered to be static in the $\mu$SR time window.

In an applied field in the superconducting phase, panel (b), best fits are obtained with a three component function (in the following we use $B_a$ and $B_c$ for the applied and critical field rather than $H_a$ and $H_c$)

\begin{multline}\label{total fitting function}
A(t)= A_{0}[f_{S}(\frac{1}{3}+\frac{2}{3}(1-\sigma_{KG}^{2}t^{2} \exp (-\frac{1}{2} \sigma_{KG}^2 t^2))
\\ + f_{N} \exp (-\frac{1}{2} \sigma_{N}^2 t^2)\cos (\gamma_{\mu}B_c t+\phi_{N})
\\+ f_{bg} \exp (-\frac{1}{2} \sigma_{bg}^2 t^2)\cos (\gamma_{\mu}B_{a}t+\phi_{bg})]
\end{multline}

\noindent
The third term, which we give the label \textquotesingle background\textquotesingle ~for the moment, is small and accounts for muons that precess in the applied field at the angular frequency $\omega = \gamma_{\mu}B_{a}$, and $\sigma_{bg}$ and $\phi_{bg}$ are the related damping and phase factor, respectively. $f_{S}=A_{S}/A_{0}$, $f_{N}=A_{N}/A_{0}$ and $f_{bg}=A_{bg}/A_{0}$ are the volume fractions related to the superconducting domains, normal domains, and the background term, respectively. $A_{0} = A_{S} + A_{N} + A_{bg}$ is the full experimental asymmetry, and was kept constant in the fitting procedure. The fit parameters at 9~mT (panel (b)) are: $f_{S} = 0.34$ (solid blue line), $f_{N} = 0.56$ and $\sigma_{N} =0.25~ \mu$s$^{-1}$ (solid green line), and $f_{bg} = 0.10$ and $\sigma_{bg} = 0.50 ~\mu$s$^{-1}$ (solid pink line). Here we have fixed $\sigma _{KG} = 0.05~\mu$s$^{-1}$. We remark that the Gaussian damping in the normal domains, $\sigma_{N} =0.25~\mu$s$^{-1}$, is larger than the value extracted from the normal state fit, see panel (c). This is not unexpected given the complicated domain patterns that can arise in the intermediate state~\cite{Huebener1979}. We will address the background term in the Discussion section.

\begin{figure}
  \centering
  \includegraphics[width= 8cm]{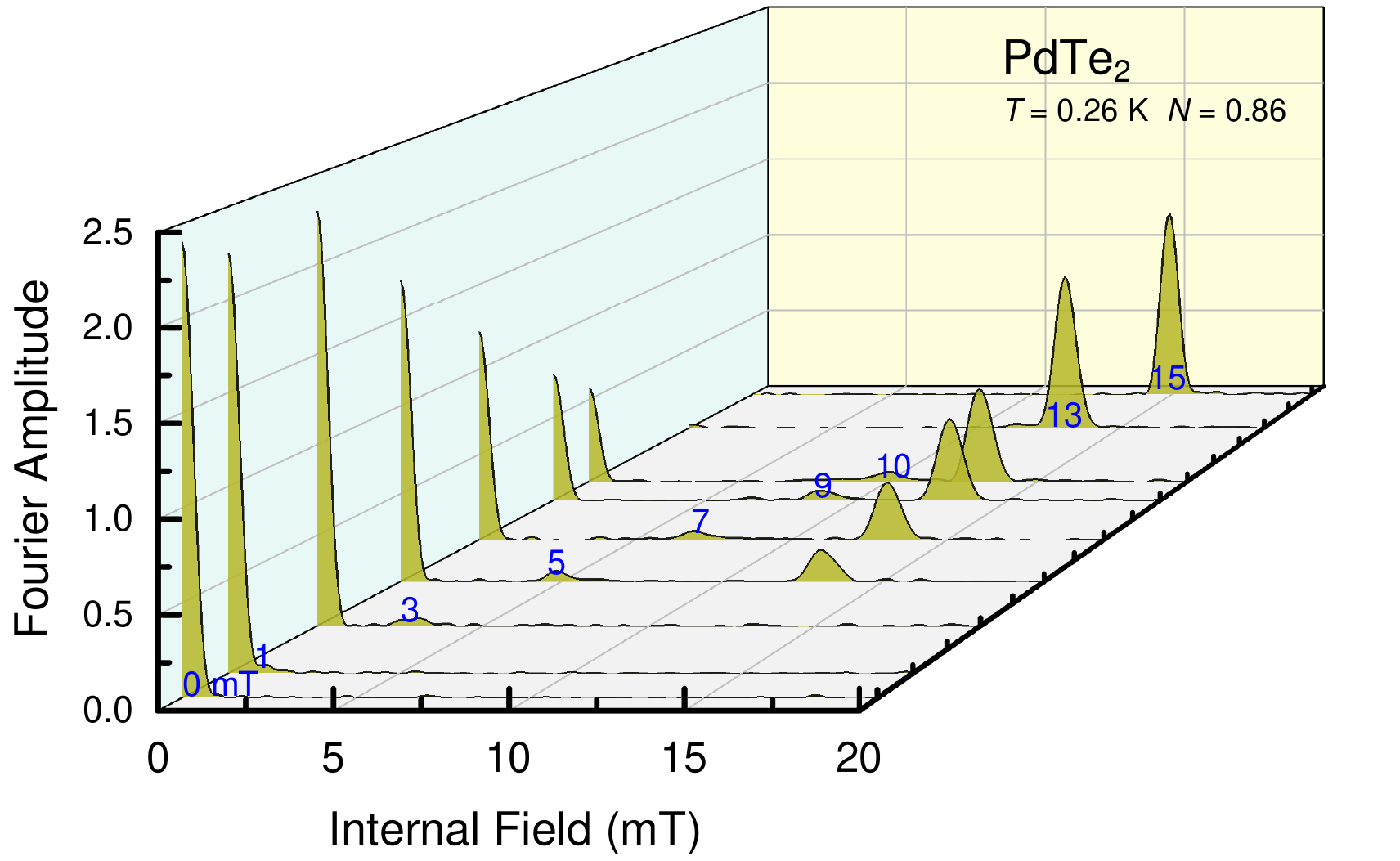}
  \caption{Magnetic field distribution in the PdTe$_2$ crystal at $T = 0.26$~K obtained by FFT for fields applied perpendicular to the sample plane. The field values are given in blue colored numbers. In the intermediate state two peaks are present at $B=0$ and at $B=B_c > B_a$. The weak intensity at $B=B_a$ signals the background contribution.}
  \label{FFT_260mK}
\end{figure}

In order to follow the evolution of the intermediate state with increasing magnetic field it is illustrative to inspect the Fast Fourier Transforms (FFT) of the $\mu$SR time spectra. The FFT amplitudes are shown in a three-dimensional (3D) plot in Fig.~\ref{FFT_260mK}. The magnetic field distributions have a sharp peak at $B=0$, which is due to the superconducting volume fraction. For $B_a = 5$~mT a second peak appears at a field $B = B_c > B_a$. This magnetic intensity is due to the normal domains. It shows the crystal is phase separated in normal and superconducting domains, as expected for the intermediate state. By further increasing the field, the peak at $B_c$ grows, while the peak at $B=0$ decreases in intensity and vanishes at $B_a = B_c$. Eventually, for $B_a = 15$~mT $> B_c =13.0$~mT, the FFT shows a peak at the applied field only. In all FFT's a low-intensity hump is visible at the applied field as well. This field distribution corresponds to the background term.

\begin{figure}
  \centering
  \includegraphics[width=7cm]{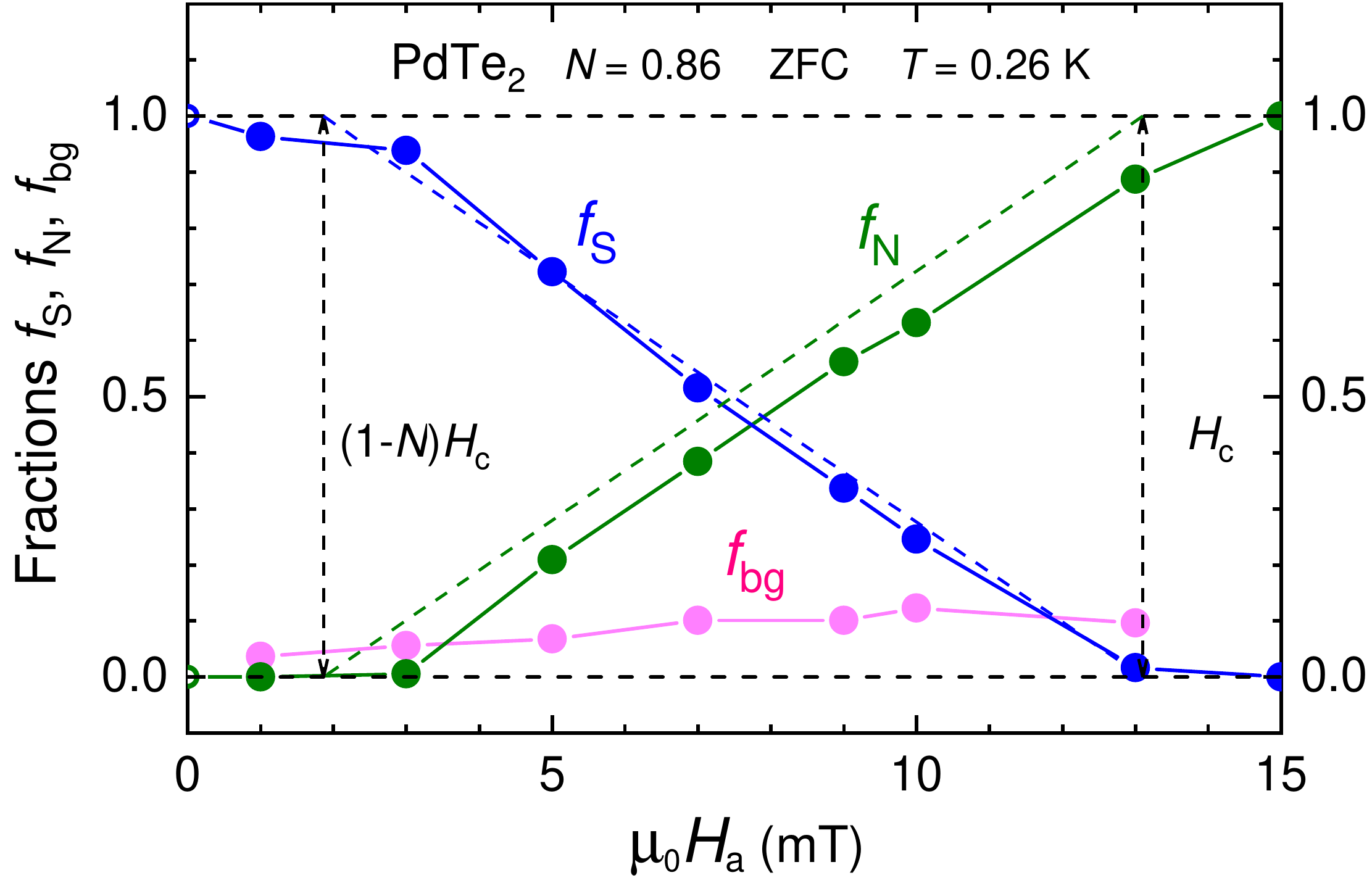}
  \caption{Field variation of the superconducting $f_S$ (blue symbols), normal $f_N$ (green symbols) and background $f_{bg}$ (pink symbols) volume fractions obtained by fitting the $\mu$SR spectra. The open symbols are ZFC at $B=0$. The vertical dashed lines at $(1-N)H_c$ and $H_c$ bound the region in which the intermediate state is expected for $N_{\perp} = 0.86$. The dashed blue and green lines show the expected linear field variation of the superconducting and normal volume fractions. The temperature is 0.26 K.}
  \label{fractions_260mK}
\end{figure}

In order to produce a quantitative analysis of the growth of the intermediate phase we have fitted the $\mu$SR spectra in applied fields to Eq.~\ref{total fitting function}, as illustrated in Fig.~\ref{spectra_260mK}(b). In Fig.~\ref{fractions_260mK} we trace the fit parameters $f_{S}$, $f_{N}$ and $f_{bg}$. In the Landau scenario the intermediate state is predicted to occur in the field range $(1-N)H_c < H_a < H_c$ and its volume fraction grows linearly $f_N (H_a) = (H_a - (1-N)H_c)/NH_c$~(Ref.~\onlinecite{Landau1937}). Overall, our results comply with the simple model, but for small fields the quasi-linear behavior does not extend all the way to $H_a = (1-N)H_c$. This points to a complex flux penetration process in weak fields. To conclude this section we remark that the value of $H_c$ at $T=0.26$~K obtained by $\mu$SR for $H_a \parallel c$, is close to the value for $H_a \parallel a$~(Ref.~\onlinecite{Leng2017}).

\subsection{Field in the plane of the disk}

\begin{figure}
  \centering
  \includegraphics[width=8cm]{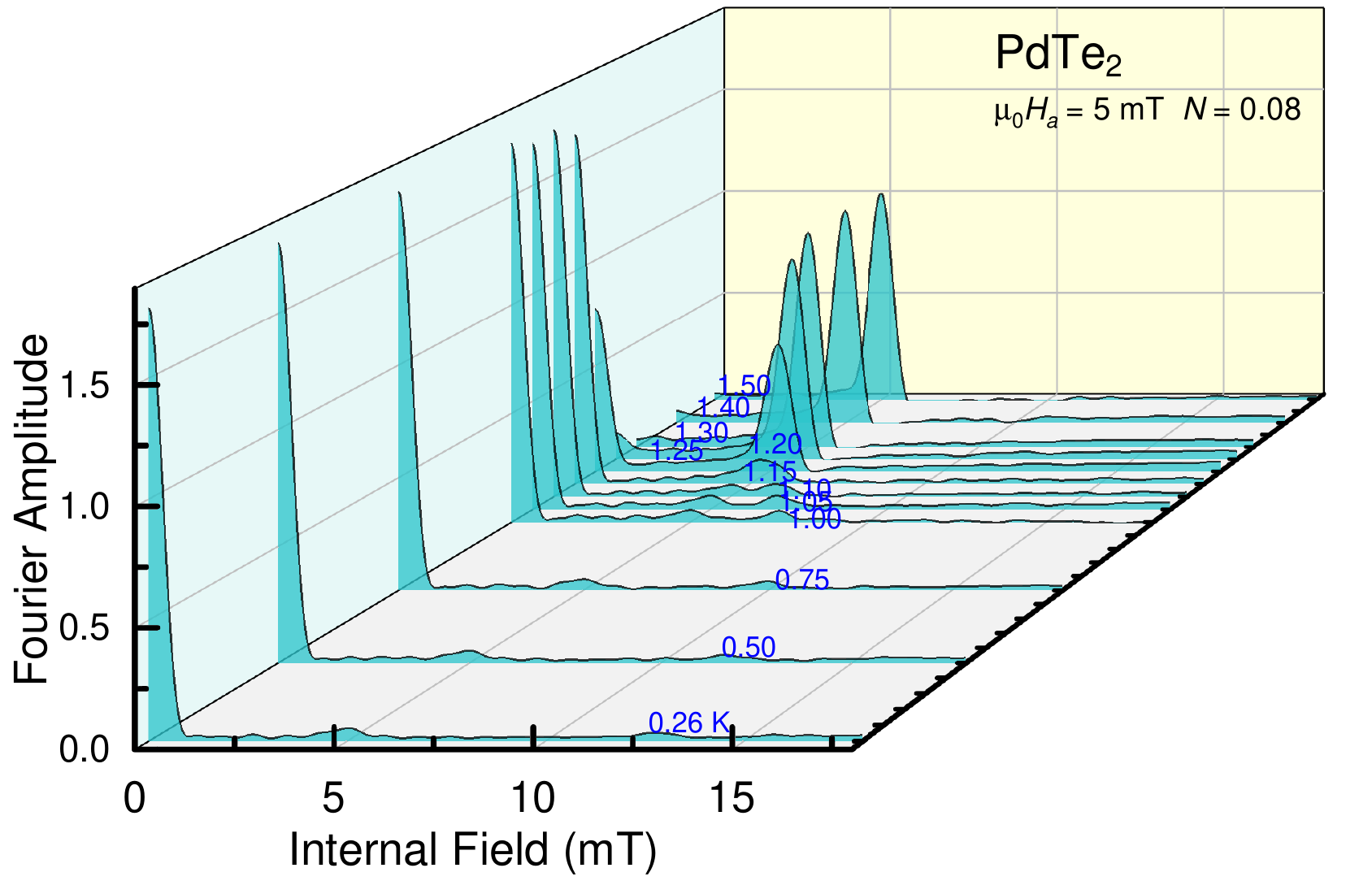}
  \caption{Magnetic field distribution in the PdTe$_2$ crystal after FC in $B_a = 5$~mT directed in the plane of the disk at different temperatures as indicated. The large peak at $B=0$ corresponds to the superconducting volume fraction. The weak intensity at $B_c (T)$ is due to a tiny part of the crystal that is already in the intermediate state at the lowest temperature (0.26~K). Upon approaching $T_c$ the whole crystal converts to the intermediate phase. The small peak that remains at $B=B_a$ signals the background contribution.}
  \label{FFT_5mT}
\end{figure}

A second set of spectra was taken after field cooling in 5~mT to a base temperature of 0.26~K, followed by stepwise heating the crystal to above $T_c$, as indicated in Fig.~\ref{procedure}(b). Here the field was applied in the plane of the disk. It is instructive to first inspect the 3D graph with the FFT's shown in Fig.~\ref{FFT_5mT}. The large peaks at $B=0$ signal the superconducting volume fraction. Surprisingly, after field cooling a tiny fraction of the crystal is in the intermediate state already, as validated by the weak magnetic intensity at $B=B_c = 13.0$~mT (at 0.26~K) $> B_a$. Upon increasing the temperature this fraction remains small up to 1.1 K. For higher temperatures the magnetic intensity at $B_c$ grows rapidly, while the peak at $B=0$ shows the opposite behavior. This shows the bulk of the crystal converts to the intermediate state. The temperature variation of $B_c$ follows the standard quadratic expression $B_c(T) = B_c(0)[1-(T/T_c)^2]$, here $B_c(0)= 13.3$~mT and $T_c = 1.53$~K. These values obtained for $H_a \parallel a^*$ are a few percent smaller than those reported in Ref.~\onlinecite{Leng2017} for $H_a \parallel a$. The low-intensity hump at $B_a = 5$~mT below $T_c$ is attributed to the background term. For $T > T_c$ the FFT peak at 5~mT is large and characterizes the paramagnetic normal-state volume of the crystal.

\begin{figure}
  \centering
  \includegraphics[width=7cm]{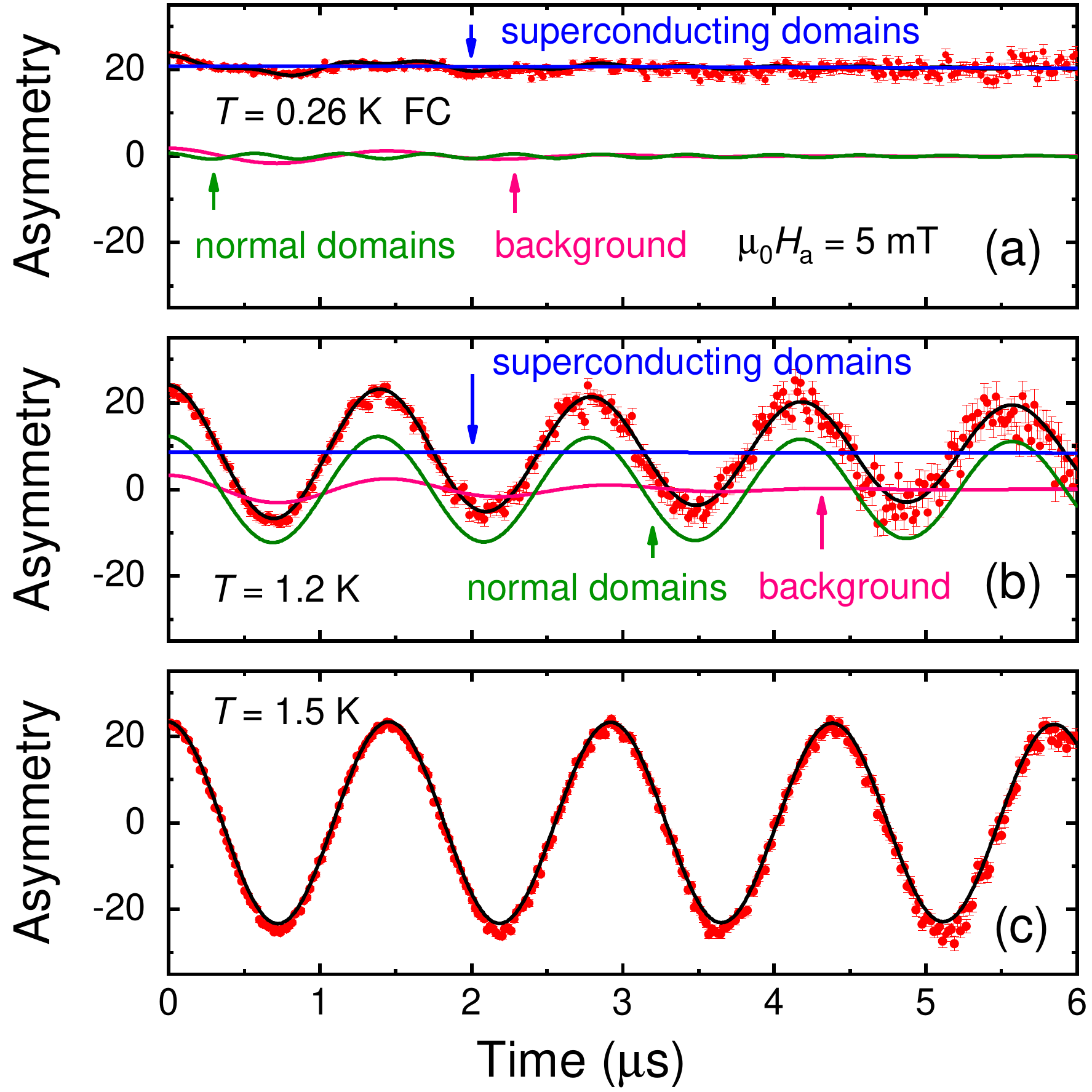}
  \caption{$\mu$SR spectra collected in a field $H_a= 5$~mT directed in the plane of the sample at 0.26~K, 1.2~K and 1.5~K. The sample is field cooled. In (a) and (b) the black line is a fit to the three component function Eq.~\ref{total fitting function}. The different components, due to superconducting domains, normal domains and background, are shown by the solid blue, green and pink lines, respectively. In (c) the black solid line is a fit to the muon depolarization function Eq.~\ref{normal}. See text for fit details.}
  \label{spectra_5mT}
\end{figure}

In Fig.~\ref{spectra_5mT} we show three typical $\mu$SR spectra from the temperature run in 5~mT together with the fit results using Eq.~\ref{normal} and \ref{total fitting function}. Here the total experimental asymmetry $A_0 = 23.3$. At 0.26~K, panel (a), the solid blue line describes the large Meissner volume, with $\sigma _{KG} = 0.03~\mu$s$^{-1}$. A tiny volume fraction with normal domains ($B_c = 13.0$~mT) shows up in the fit as well (solid green line), which indicates a tiny part of the crystal is in the intermediate state. At 1.2~K, panel (b), the normal state domains occupy about half of the crystal's volume. This is shown as the solid green line, which is the Gaussian damped oscillatory component with $\sigma_{N} =0.08~\mu$s$^{-1}$. At 1.5~K, panel (c), the crystal is the normal state. The data are well fitted by Eq.~\ref{normal} with the small relaxation rate $\sigma _{N} = 0.04~\mu$s$^{-1}$ (black solid line).

\begin{figure}
  \centering
  \includegraphics[width=7cm]{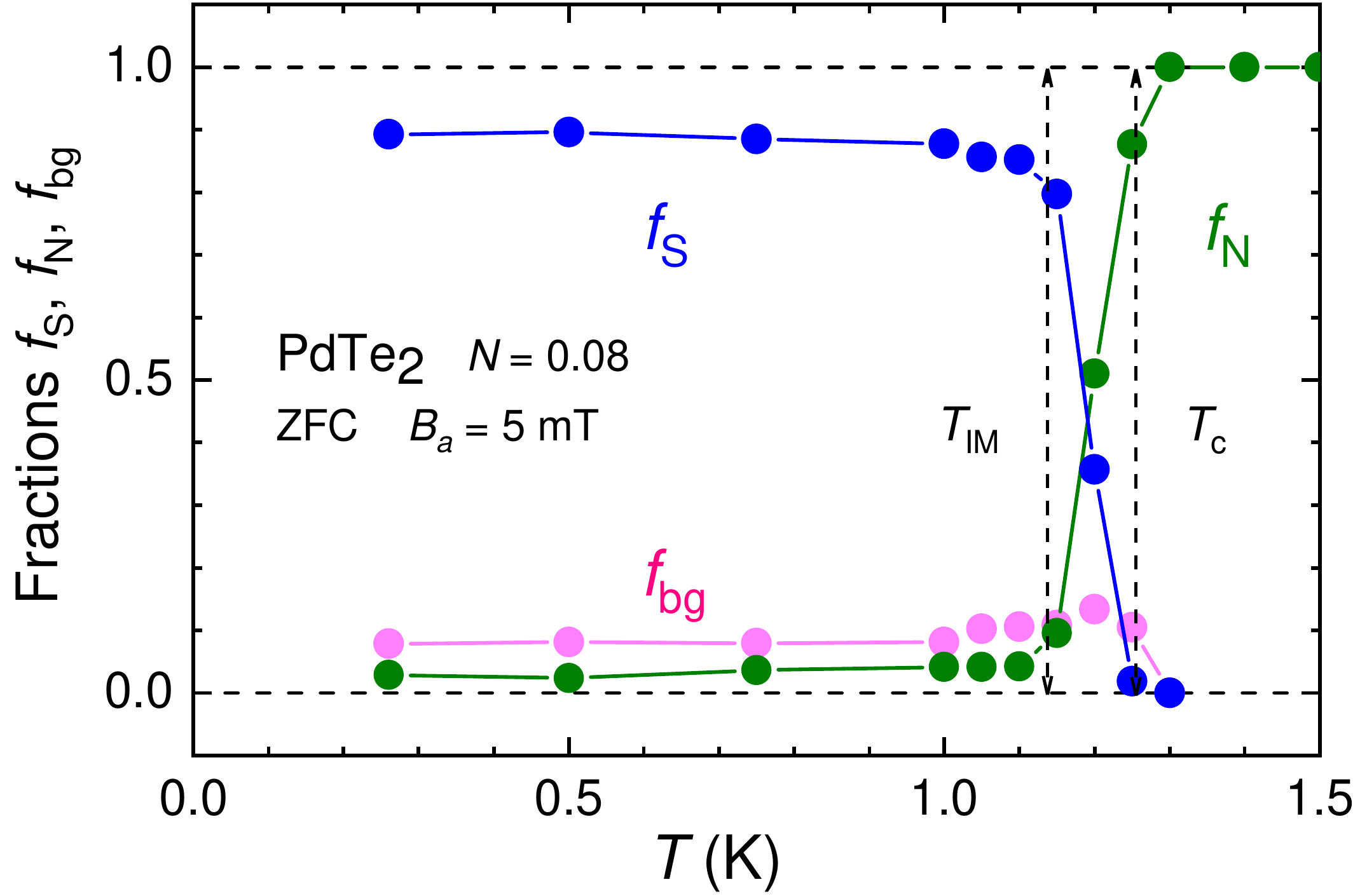}
  \caption{Temperature variation of the superconducting $f_S$ (blue symbols), normal $f_N$ (green symbols) and background $f_{bg}$ (pink symbols) volume fractions obtained by fitting the $\mu$SR spectra using Eq.~\ref{total fitting function} (FC 5~mT directed in the plane of the disk). The vertical dashed lines at $T_{IM}$ and $T_c$ bound the region in which the intermediate state in the bulk of the crystal is found. }
  \label{fractions_5mT}
\end{figure}

In Fig.~\ref{fractions_5mT} we trace the different volume fractions as a function of temperature obtained by fitting all the spectra. Clearly, during field cooling some flux remains trapped in the crystal, resulting in a superconducting volume fraction $f_S \simeq 0.90$. The tiny volume fraction with normal domains (internal field $B_c$) does not vary with temperature below $\sim 1.1$~K and equals $f_N \simeq 0.02$. This implies that the Meissner fraction in this bulky sample occupies $\sim$ 90\% of its volume, which may be compared with the value of 93\% obtained for a small crystal measured via dc-magnetization~\cite{Leng2017}. The presence of a tiny intermediate state fraction is most likely related to the edges of the crystal that may result locally in  a large demagnetization factor. Upon raising the temperature the bulk of the crystal transforms to the intermediate state above $\sim 1.1$~K. While $f_N$ grows steeply, $f_S$ decreases. In Fig.~\ref{fractions_5mT} we have indicated the borders of the intermediate phase by the vertical dashed lines at $T_{IM} = 1.14$~K and $T_c = 1.25$~K. The temperature at which the transformation starts is lower than can be expected on the basis of the demagnetization factor $N=0.08$. This indicates a larger, effective demagnetization factor $N_{eff}$. With $T_{IM} = 1.14$~K, we calculate $N_{eff} = 0.16$.

\section{DISCUSSION}

The most important conclusion that can be drawn from our $\mu$SR experiments is that the bulk of our PdTe$_2$ crystal exhibits type-I superconductivity. Solid evidence for this is provided by the detection of the intermediate phase. Here we use the muon as a local probe of the bulk on the microscopic level. It is of interest to provide a lower bound of the crystal volume that is occupied by type-I superconductivity. It cannot simply be taken equal to the ZFC Meissner volume, $f_S = 1$, deduced from Fig.~\ref{spectra_260mK}(a), because muons stopping in a (tiny) non-superconducting part of the crystal will experience a similar Gaussian Kubo-Toyabe depolarization as muons in the superconducting part, and thus cannot be distinguished. However, an estimate can be made by considering the intermediate phase fraction, $f_{IM} = f_S+f_N$. From the data in Fig.~\ref{fractions_260mK} a lower bound for $f_{IM}$ can be obtained by linearly extrapolating $f_N (H_a)$ to $H_c$, where $f_S = 0$. We find $f_N = f_{IM} = 0.92$. On the same grounds, $f_S = f_{IM} = 0.94$ at the start of the linear growth of $f_N$. This tells us type-I superconductivity occupies at least 92\% of the crystal's volume.

Next we address the background term, that results in the remaining volume fraction (5-10\%) due to the third component in Eq.~\ref{total fitting function}, \textit{i.e.} muons that precess at the frequency of the applied field. Since the muons and decay positrons events are collected in the so-called VETO mode, the contribution from positrons arising from muons that do not stop in the sample will be small. Besides, the damping rate (\textit{e.g.} $\sigma_{bg} = 0.50 ~\mu$s$^{-1}$ for the spectrum in Fig.~\ref{spectra_260mK}(b)) is too large to stem from the usual background components, such as the sample holder and cryostat, and indicates a local broad field distribution. This hints at an intrinsic source of inhomogeneities related to type-I superconductivity. In general the penetration or expulsion of flux in a type-I superconductor is a complicated process, and the domain pattern in the intermediate state can be diverse and complex~\cite{Huebener1979}. Moreover, the demagnetization factor in the crystal is not uniform, especially near the edges. This brings about additional internal field inhomogeneities, as illustrated by the tiny intermediate state fraction observed with the field in the plane of the disk.

Another aspect is that the superconducting and normal domains in the intermediate state are separated by domain walls. The width of the domain wall~\cite{Huebener1979} is of the order $\delta \sim \xi - \lambda \approx 1.3~\mu$m~\cite{Salis2018}. In the ideal case of a laminar domain pattern an estimate for the volume fraction of the domain walls is $f_{DW} = 2 \delta / a$, where the periodicity length $a = (d \delta / f(\tilde{h}))^{1/2}$, see Ref.~\onlinecite{Huebener1979}. Here $d = 0.65$~mm is the sample thickness and $f(\tilde{h})$ a numerical function with $\tilde{h} = H_a/H_c$. For an applied field of typically 5~mT (Fig.~\ref{spectra_5mT}), $\tilde{h} = 0.38$ and $f(\tilde{h}) = 0.022$. Consequently, $f_{DW} \approx $~1.4\%. It is not surprising that this value is considerably smaller than $f_{bg}$ measured, because the domain patterns in our crystal will be complex, and concurrently the domain walls broad. We therefore argue that muons stopping in domain walls can largely account for the background term. Besides, muons stopping in regions where the magnetic field is pinned or trapped at defects during flux penetration or expulsion will contribute as well. Considering that the background term can be accounted for by these sources of $\mu^+$-spin depolarization, the data do not rule out that the type-I superconducting fraction in our crystal is close to 100\%.

On the other hand, the possibility that a minute fraction of the crystal exhibits type-II superconductivity cannot be completely dismissed. In a type-II superconductor the local field in the vortex phase is close to the applied field and thus its field distribution could contribute to $f_{bg}$. Local type-II behavior could possibly originate from a pronounced deviation of the 1:2 stoichiometry. We recall, however, that the EDX spectra show a uniform 1:2 composition within the experimental resolution of 0.5\%. A mixed type-I and type-II behavior has been evoked to explain the STM/STS and PCS spectra, measured at the surface of PdTe$_2$~\cite{Sirohi2019,Le2019}. Here it is proposed that the electron mean free path, $\ell$, is locally reduced, which results in $\kappa > 1/\sqrt 2$. We remark, evidence for flux quantization and a vortex lattice required for type-II superconductivity has not been produced. STM/STS and PCS are surface sensitive probes, and thus possibly the mixed behavior is a property of the crystal's surface only. But this in turn is difficult to reconcile with the resulting field of the vortex that has to penetrate the bulk. It is tempting to speculate that these unusual surface effects, as well as the superconductivity of the surface sheath~\cite{Leng2017}, are related to the Dirac type-II character that involves topological surface states. This warrants a continuing investigation of PdTe$_2$. Superconductivity of the surface sheath~\cite{Leng2017} has been detected by magnetic susceptibility in small ac-driving fields only, and could not be probed in the present $\mu$SR experiments, which employs dc-fields. In order to obtain access to the surface properties Low Energy Muons (LEM) form an excellent tool.  Here the energy of the muons can be tuned such that they localize in the surface layer of the crystal. However, at the moment this $\mu$SR technique is restricted to temperatures above 2~K only.

\section{SUMMARY}

We have investigated the superconducting phase of PdTe$_2$ ($T_c = 1.6$~K) by transverse field muon spin rotation experiments. $\mu$SR spectra were taken on a thin disk-like crystal in two configurations: with the field perpendicular to the plane of the disk ($N_{\perp}=0.86$) and with the field in the plane of the disk ($N_{\parallel}=0.08$). The $H-T$ phase diagram was scanned as a function of temperature and applied field. The $\mu$SR spectra have been analysed with a three component muon depolarization function, accounting for the superconducting domains, the normal domains and a background term. In the superconducting phase normal domains are found in which the local field is always equal to $B_c$ and larger than the applied field. This is the hall mark of the intermediate phase in a type-I superconductor. The background term is predominantly attributed to muons stopping in the superconducting-normal domain walls. In conclusion, our $\mu$SR study provides solid evidence for type-I behavior in the bulk of the PdTe$_2$ crystal.

\vspace{5mm}

\section{acknowledgements}
H.L. acknowledges the Chinese Scholarship Council for Grant No.~201604910855. This work was part of the research program on Topological Insulators funded by FOM (Dutch Foundation for Fundamental Research on Matter).

\bibliography{PdTe2_musr}

\begin{thebibliography}{44}%
\makeatletter
\providecommand \@ifxundefined [1]{%
 \@ifx{#1\undefined}
}%
\providecommand \@ifnum [1]{%
 \ifnum #1\expandafter \@firstoftwo
 \else \expandafter \@secondoftwo
 \fi
}%
\providecommand \@ifx [1]{%
 \ifx #1\expandafter \@firstoftwo
 \else \expandafter \@secondoftwo
 \fi
}%
\providecommand \natexlab [1]{#1}%
\providecommand \enquote  [1]{``#1''}%
\providecommand \bibnamefont  [1]{#1}%
\providecommand \bibfnamefont [1]{#1}%
\providecommand \citenamefont [1]{#1}%
\providecommand \href@noop [0]{\@secondoftwo}%
\providecommand \href [0]{\begingroup \@sanitize@url \@href}%
\providecommand \@href[1]{\@@startlink{#1}\@@href}%
\providecommand \@@href[1]{\endgroup#1\@@endlink}%
\providecommand \@sanitize@url [0]{\catcode `\\12\catcode `\$12\catcode
  `\&12\catcode `\#12\catcode `\^12\catcode `\_12\catcode `\%12\relax}%
\providecommand \@@startlink[1]{}%
\providecommand \@@endlink[0]{}%
\providecommand \url  [0]{\begingroup\@sanitize@url \@url }%
\providecommand \@url [1]{\endgroup\@href {#1}{\urlprefix }}%
\providecommand \urlprefix  [0]{URL }%
\providecommand \Eprint [0]{\href }%
\providecommand \doibase [0]{http://dx.doi.org/}%
\providecommand \selectlanguage [0]{\@gobble}%
\providecommand \bibinfo  [0]{\@secondoftwo}%
\providecommand \bibfield  [0]{\@secondoftwo}%
\providecommand \translation [1]{[#1]}%
\providecommand \BibitemOpen [0]{}%
\providecommand \bibitemStop [0]{}%
\providecommand \bibitemNoStop [0]{.\EOS\space}%
\providecommand \EOS [0]{\spacefactor3000\relax}%
\providecommand \BibitemShut  [1]{\csname bibitem#1\endcsname}%
\let\auto@bib@innerbib\@empty
\bibitem [{\citenamefont {Soluyanov}\ \emph {et~al.}(2015)\citenamefont
  {Soluyanov}, \citenamefont {Gresch}, \citenamefont {Wang}, \citenamefont
  {Wu}, \citenamefont {Troyer}, \citenamefont {Dai},\ and\ \citenamefont
  {Bernevig}}]{Soluyanov2015}%
  \BibitemOpen
  \bibfield  {author} {\bibinfo {author} {\bibfnamefont {A.~A.}\ \bibnamefont
  {Soluyanov}}, \bibinfo {author} {\bibfnamefont {D.}~\bibnamefont {Gresch}},
  \bibinfo {author} {\bibfnamefont {Z.}~\bibnamefont {Wang}}, \bibinfo {author}
  {\bibfnamefont {Q.}~\bibnamefont {Wu}}, \bibinfo {author} {\bibfnamefont
  {M.}~\bibnamefont {Troyer}}, \bibinfo {author} {\bibfnamefont
  {X.}~\bibnamefont {Dai}}, \ and\ \bibinfo {author} {\bibfnamefont
  {B.}~\bibnamefont {Bernevig}},\ }\href@noop {} {\bibfield  {journal}
  {\bibinfo  {journal} {Nature}\ }\textbf {\bibinfo {volume} {257}},\ \bibinfo
  {pages} {495} (\bibinfo {year} {2015})}\BibitemShut {NoStop}%
\bibitem [{\citenamefont {Huang}\ \emph {et~al.}(2016)\citenamefont {Huang},
  \citenamefont {Zhou},\ and\ \citenamefont {Duan}}]{Huang2016}%
  \BibitemOpen
  \bibfield  {author} {\bibinfo {author} {\bibfnamefont {H.}~\bibnamefont
  {Huang}}, \bibinfo {author} {\bibfnamefont {S.}~\bibnamefont {Zhou}}, \ and\
  \bibinfo {author} {\bibfnamefont {W.}~\bibnamefont {Duan}},\ }\href {\doibase
  10.1103/PhysRevB.94.121117} {\bibfield  {journal} {\bibinfo  {journal} {Phys.
  Rev. B}\ }\textbf {\bibinfo {volume} {94}},\ \bibinfo {pages} {121117}
  (\bibinfo {year} {2016})}\BibitemShut {NoStop}%
\bibitem [{\citenamefont {Yan}\ \emph {et~al.}(2017)\citenamefont {Yan},
  \citenamefont {Huang}, \citenamefont {Zhang}, \citenamefont {Wang},
  \citenamefont {Yao}, \citenamefont {Deng}, \citenamefont {Wan}, \citenamefont
  {Zhang}, \citenamefont {Arita}, \citenamefont {Yang}, \citenamefont {Sun},
  \citenamefont {Yao}, \citenamefont {Wu}, \citenamefont {Fan}, \citenamefont
  {Duan},\ and\ \citenamefont {Zhou}}]{Yan2017}%
  \BibitemOpen
  \bibfield  {author} {\bibinfo {author} {\bibfnamefont {M.}~\bibnamefont
  {Yan}}, \bibinfo {author} {\bibfnamefont {H.}~\bibnamefont {Huang}}, \bibinfo
  {author} {\bibfnamefont {K.}~\bibnamefont {Zhang}}, \bibinfo {author}
  {\bibfnamefont {E.}~\bibnamefont {Wang}}, \bibinfo {author} {\bibfnamefont
  {W.}~\bibnamefont {Yao}}, \bibinfo {author} {\bibfnamefont {K.}~\bibnamefont
  {Deng}}, \bibinfo {author} {\bibfnamefont {G.}~\bibnamefont {Wan}}, \bibinfo
  {author} {\bibfnamefont {H.}~\bibnamefont {Zhang}}, \bibinfo {author}
  {\bibfnamefont {M.}~\bibnamefont {Arita}}, \bibinfo {author} {\bibfnamefont
  {H.}~\bibnamefont {Yang}}, \bibinfo {author} {\bibfnamefont {Z.}~\bibnamefont
  {Sun}}, \bibinfo {author} {\bibfnamefont {H.}~\bibnamefont {Yao}}, \bibinfo
  {author} {\bibfnamefont {Y.}~\bibnamefont {Wu}}, \bibinfo {author}
  {\bibfnamefont {S.}~\bibnamefont {Fan}}, \bibinfo {author} {\bibfnamefont
  {W.}~\bibnamefont {Duan}}, \ and\ \bibinfo {author} {\bibfnamefont
  {S.}~\bibnamefont {Zhou}},\ }\href@noop {} {\bibfield  {journal} {\bibinfo
  {journal} {Nature Comm.}\ }\textbf {\bibinfo {volume} {8}},\ \bibinfo {pages}
  {257} (\bibinfo {year} {2017})}\BibitemShut {NoStop}%
\bibitem [{\citenamefont {Bahramy}\ \emph {et~al.}(2018)\citenamefont
  {Bahramy}, \citenamefont {Clark}, \citenamefont {Yang}, \citenamefont {Feng},
  \citenamefont {Bawden}, \citenamefont {Riley}, \citenamefont {Markovic},
  \citenamefont {Mazzola}, \citenamefont {Sunko}, \citenamefont {Biswas},
  \citenamefont {Cooil}, \citenamefont {Jorge}, \citenamefont {Wells},
  \citenamefont {Leandersson}, \citenamefont {Balasubramanian}, \citenamefont
  {Fujii}, \citenamefont {Vobornik}, \citenamefont {Rault}, \citenamefont
  {Kim}, \citenamefont {Hoesch}, \citenamefont {Okawa}, \citenamefont
  {Asakawa}, \citenamefont {Sasagawa}, \citenamefont {Eknapakul}, \citenamefont
  {Meevasana},\ and\ \citenamefont {King}}]{Bahramy2018}%
  \BibitemOpen
  \bibfield  {author} {\bibinfo {author} {\bibfnamefont {M.~S.}\ \bibnamefont
  {Bahramy}}, \bibinfo {author} {\bibfnamefont {O.~J.}\ \bibnamefont {Clark}},
  \bibinfo {author} {\bibfnamefont {B.-J.}\ \bibnamefont {Yang}}, \bibinfo
  {author} {\bibfnamefont {J.}~\bibnamefont {Feng}}, \bibinfo {author}
  {\bibfnamefont {L.}~\bibnamefont {Bawden}}, \bibinfo {author} {\bibfnamefont
  {J.~M.}\ \bibnamefont {Riley}}, \bibinfo {author} {\bibfnamefont
  {I.}~\bibnamefont {Markovic}}, \bibinfo {author} {\bibfnamefont
  {F.}~\bibnamefont {Mazzola}}, \bibinfo {author} {\bibfnamefont
  {V.}~\bibnamefont {Sunko}}, \bibinfo {author} {\bibfnamefont
  {D.}~\bibnamefont {Biswas}}, \bibinfo {author} {\bibfnamefont {S.~P.}\
  \bibnamefont {Cooil}}, \bibinfo {author} {\bibfnamefont {M.}~\bibnamefont
  {Jorge}}, \bibinfo {author} {\bibfnamefont {J.~W.}\ \bibnamefont {Wells}},
  \bibinfo {author} {\bibfnamefont {M.}~\bibnamefont {Leandersson}}, \bibinfo
  {author} {\bibfnamefont {T.}~\bibnamefont {Balasubramanian}}, \bibinfo
  {author} {\bibfnamefont {J.}~\bibnamefont {Fujii}}, \bibinfo {author}
  {\bibfnamefont {I.}~\bibnamefont {Vobornik}}, \bibinfo {author}
  {\bibfnamefont {J.~E.}\ \bibnamefont {Rault}}, \bibinfo {author}
  {\bibfnamefont {T.~K.}\ \bibnamefont {Kim}}, \bibinfo {author} {\bibfnamefont
  {M.}~\bibnamefont {Hoesch}}, \bibinfo {author} {\bibfnamefont
  {K.}~\bibnamefont {Okawa}}, \bibinfo {author} {\bibfnamefont
  {M.}~\bibnamefont {Asakawa}}, \bibinfo {author} {\bibfnamefont
  {T.}~\bibnamefont {Sasagawa}}, \bibinfo {author} {\bibfnamefont
  {T.}~\bibnamefont {Eknapakul}}, \bibinfo {author} {\bibfnamefont
  {W.}~\bibnamefont {Meevasana}}, \ and\ \bibinfo {author} {\bibfnamefont
  {P.~D.~C.}\ \bibnamefont {King}},\ }\href {\doibase 10.1038/nmat5031}
  {\bibfield  {journal} {\bibinfo  {journal} {Nature Mat.}\ }\textbf {\bibinfo
  {volume} {17}},\ \bibinfo {pages} {21} (\bibinfo {year} {2018})}\BibitemShut
  {NoStop}%
\bibitem [{\citenamefont {Yan}\ \emph {et~al.}(2015)\citenamefont {Yan},
  \citenamefont {Jian-Zhou}, \citenamefont {Li}, \citenamefont {Cheng-Tian},
  \citenamefont {Ai-Ji}, \citenamefont {Cheng}, \citenamefont {Ying},
  \citenamefont {Yu}, \citenamefont {Shao-Long}, \citenamefont {Lin},
  \citenamefont {Guo-Dong}, \citenamefont {Xiao-Li}, \citenamefont {Jun},
  \citenamefont {Chuang-Tian}, \citenamefont {Zu-Yan}, \citenamefont
  {Hong-Ming}, \citenamefont {Xi}, \citenamefont {Zhong},\ and\ \citenamefont
  {Xing-Jiang}}]{Liu2015a}%
  \BibitemOpen
  \bibfield  {author} {\bibinfo {author} {\bibfnamefont {L.}~\bibnamefont
  {Yan}}, \bibinfo {author} {\bibfnamefont {Z.}~\bibnamefont {Jian-Zhou}},
  \bibinfo {author} {\bibfnamefont {Y.}~\bibnamefont {Li}}, \bibinfo {author}
  {\bibfnamefont {L.}~\bibnamefont {Cheng-Tian}}, \bibinfo {author}
  {\bibfnamefont {L.}~\bibnamefont {Ai-Ji}}, \bibinfo {author} {\bibfnamefont
  {H.}~\bibnamefont {Cheng}}, \bibinfo {author} {\bibfnamefont
  {D.}~\bibnamefont {Ying}}, \bibinfo {author} {\bibfnamefont {X.}~\bibnamefont
  {Yu}}, \bibinfo {author} {\bibfnamefont {H.}~\bibnamefont {Shao-Long}},
  \bibinfo {author} {\bibfnamefont {Z.}~\bibnamefont {Lin}}, \bibinfo {author}
  {\bibfnamefont {L.}~\bibnamefont {Guo-Dong}}, \bibinfo {author}
  {\bibfnamefont {D.}~\bibnamefont {Xiao-Li}}, \bibinfo {author} {\bibfnamefont
  {Z.}~\bibnamefont {Jun}}, \bibinfo {author} {\bibfnamefont {C.}~\bibnamefont
  {Chuang-Tian}}, \bibinfo {author} {\bibfnamefont {X.}~\bibnamefont {Zu-Yan}},
  \bibinfo {author} {\bibfnamefont {W.}~\bibnamefont {Hong-Ming}}, \bibinfo
  {author} {\bibfnamefont {D.}~\bibnamefont {Xi}}, \bibinfo {author}
  {\bibfnamefont {F.}~\bibnamefont {Zhong}}, \ and\ \bibinfo {author}
  {\bibfnamefont {Z.}~\bibnamefont {Xing-Jiang}},\ }\href {\doibase
  10.1088/0256-307X/32/6/067303} {\bibfield  {journal} {\bibinfo  {journal}
  {Chin. Phys. Lett.}\ }\textbf {\bibinfo {volume} {32}},\ \bibinfo {eid}
  {067303} (\bibinfo {year} {2015})}\BibitemShut {NoStop}%
\bibitem [{\citenamefont {Fei}\ \emph {et~al.}(2017)\citenamefont {Fei},
  \citenamefont {Bo}, \citenamefont {Wang}, \citenamefont {Wu}, \citenamefont
  {Jiang}, \citenamefont {Fu}, \citenamefont {Gao}, \citenamefont {Zheng},
  \citenamefont {Chen}, \citenamefont {Wang}, \citenamefont {Bu}, \citenamefont
  {Song}, \citenamefont {Wan}, \citenamefont {Wang},\ and\ \citenamefont
  {Wang}}]{Fei2017}%
  \BibitemOpen
  \bibfield  {author} {\bibinfo {author} {\bibfnamefont {F.}~\bibnamefont
  {Fei}}, \bibinfo {author} {\bibfnamefont {X.}~\bibnamefont {Bo}}, \bibinfo
  {author} {\bibfnamefont {R.}~\bibnamefont {Wang}}, \bibinfo {author}
  {\bibfnamefont {B.}~\bibnamefont {Wu}}, \bibinfo {author} {\bibfnamefont
  {J.}~\bibnamefont {Jiang}}, \bibinfo {author} {\bibfnamefont
  {D.}~\bibnamefont {Fu}}, \bibinfo {author} {\bibfnamefont {M.}~\bibnamefont
  {Gao}}, \bibinfo {author} {\bibfnamefont {H.}~\bibnamefont {Zheng}}, \bibinfo
  {author} {\bibfnamefont {Y.}~\bibnamefont {Chen}}, \bibinfo {author}
  {\bibfnamefont {X.}~\bibnamefont {Wang}}, \bibinfo {author} {\bibfnamefont
  {H.}~\bibnamefont {Bu}}, \bibinfo {author} {\bibfnamefont {F.}~\bibnamefont
  {Song}}, \bibinfo {author} {\bibfnamefont {X.}~\bibnamefont {Wan}}, \bibinfo
  {author} {\bibfnamefont {B.}~\bibnamefont {Wang}}, \ and\ \bibinfo {author}
  {\bibfnamefont {G.}~\bibnamefont {Wang}},\ }\href {\doibase
  10.1103/PhysRevB.96.041201} {\bibfield  {journal} {\bibinfo  {journal} {Phys.
  Rev. B}\ }\textbf {\bibinfo {volume} {96}},\ \bibinfo {pages} {041201}
  (\bibinfo {year} {2017})}\BibitemShut {NoStop}%
\bibitem [{\citenamefont {Noh}\ \emph {et~al.}(2017)\citenamefont {Noh},
  \citenamefont {Jeong}, \citenamefont {Cho}, \citenamefont {Kim},
  \citenamefont {Min},\ and\ \citenamefont {Park}}]{Noh2017}%
  \BibitemOpen
  \bibfield  {author} {\bibinfo {author} {\bibfnamefont {H.-J.}\ \bibnamefont
  {Noh}}, \bibinfo {author} {\bibfnamefont {J.}~\bibnamefont {Jeong}}, \bibinfo
  {author} {\bibfnamefont {E.-J.}\ \bibnamefont {Cho}}, \bibinfo {author}
  {\bibfnamefont {K.}~\bibnamefont {Kim}}, \bibinfo {author} {\bibfnamefont
  {B.~I.}\ \bibnamefont {Min}}, \ and\ \bibinfo {author} {\bibfnamefont
  {B.-G.}\ \bibnamefont {Park}},\ }\href {\doibase
  10.1103/PhysRevLett.119.016401} {\bibfield  {journal} {\bibinfo  {journal}
  {Phys. Rev. Lett.}\ }\textbf {\bibinfo {volume} {119}},\ \bibinfo {pages}
  {016401} (\bibinfo {year} {2017})}\BibitemShut {NoStop}%
\bibitem [{\citenamefont {Clark}\ \emph {et~al.}(2018)\citenamefont {Clark},
  \citenamefont {Neat}, \citenamefont {Okawa}, \citenamefont {Bawden},
  \citenamefont {Markovi\ifmmode~\acute{c}\else \'{c}\fi{}}, \citenamefont
  {Mazzola}, \citenamefont {Feng}, \citenamefont {Sunko}, \citenamefont
  {Riley}, \citenamefont {Meevasana}, \citenamefont {Fujii}, \citenamefont
  {Vobornik}, \citenamefont {Kim}, \citenamefont {Hoesch}, \citenamefont
  {Sasagawa}, \citenamefont {Wahl}, \citenamefont {Bahramy},\ and\
  \citenamefont {King}}]{Clark2017}%
  \BibitemOpen
  \bibfield  {author} {\bibinfo {author} {\bibfnamefont {O.~J.}\ \bibnamefont
  {Clark}}, \bibinfo {author} {\bibfnamefont {M.~J.}\ \bibnamefont {Neat}},
  \bibinfo {author} {\bibfnamefont {K.}~\bibnamefont {Okawa}}, \bibinfo
  {author} {\bibfnamefont {L.}~\bibnamefont {Bawden}}, \bibinfo {author}
  {\bibfnamefont {I.}~\bibnamefont {Markovi\ifmmode~\acute{c}\else
  \'{c}\fi{}}}, \bibinfo {author} {\bibfnamefont {F.}~\bibnamefont {Mazzola}},
  \bibinfo {author} {\bibfnamefont {J.}~\bibnamefont {Feng}}, \bibinfo {author}
  {\bibfnamefont {V.}~\bibnamefont {Sunko}}, \bibinfo {author} {\bibfnamefont
  {J.~M.}\ \bibnamefont {Riley}}, \bibinfo {author} {\bibfnamefont
  {W.}~\bibnamefont {Meevasana}}, \bibinfo {author} {\bibfnamefont
  {J.}~\bibnamefont {Fujii}}, \bibinfo {author} {\bibfnamefont
  {I.}~\bibnamefont {Vobornik}}, \bibinfo {author} {\bibfnamefont {T.~K.}\
  \bibnamefont {Kim}}, \bibinfo {author} {\bibfnamefont {M.}~\bibnamefont
  {Hoesch}}, \bibinfo {author} {\bibfnamefont {T.}~\bibnamefont {Sasagawa}},
  \bibinfo {author} {\bibfnamefont {P.}~\bibnamefont {Wahl}}, \bibinfo {author}
  {\bibfnamefont {M.~S.}\ \bibnamefont {Bahramy}}, \ and\ \bibinfo {author}
  {\bibfnamefont {P.~D.~C.}\ \bibnamefont {King}},\ }\href {\doibase
  10.1103/PhysRevLett.120.156401} {\bibfield  {journal} {\bibinfo  {journal}
  {Phys. Rev. Lett.}\ }\textbf {\bibinfo {volume} {120}},\ \bibinfo {pages}
  {156401} (\bibinfo {year} {2018})}\BibitemShut {NoStop}%
\bibitem [{\citenamefont {Guggenheim}\ \emph {et~al.}(1961)\citenamefont
  {Guggenheim}, \citenamefont {Hulliger},\ and\ \citenamefont
  {M\"{u}ller}}]{Guggenheim1961}%
  \BibitemOpen
  \bibfield  {author} {\bibinfo {author} {\bibfnamefont {J.}~\bibnamefont
  {Guggenheim}}, \bibinfo {author} {\bibfnamefont {F.}~\bibnamefont
  {Hulliger}}, \ and\ \bibinfo {author} {\bibfnamefont {J.}~\bibnamefont
  {M\"{u}ller}},\ }\href@noop {} {\bibfield  {journal} {\bibinfo  {journal}
  {Helv. Phys. Acta}\ }\textbf {\bibinfo {volume} {34}},\ \bibinfo {pages}
  {408} (\bibinfo {year} {1961})}\BibitemShut {NoStop}%
\bibitem [{\citenamefont {Leng}\ \emph {et~al.}(2017)\citenamefont {Leng},
  \citenamefont {Paulsen}, \citenamefont {Huang},\ and\ \citenamefont
  {de~Visser}}]{Leng2017}%
  \BibitemOpen
  \bibfield  {author} {\bibinfo {author} {\bibfnamefont {H.}~\bibnamefont
  {Leng}}, \bibinfo {author} {\bibfnamefont {C.}~\bibnamefont {Paulsen}},
  \bibinfo {author} {\bibfnamefont {Y.~K.}\ \bibnamefont {Huang}}, \ and\
  \bibinfo {author} {\bibfnamefont {A.}~\bibnamefont {de~Visser}},\ }\href
  {\doibase 10.1103/PhysRevB.96.220506} {\bibfield  {journal} {\bibinfo
  {journal} {Phys. Rev. B}\ }\textbf {\bibinfo {volume} {96}},\ \bibinfo
  {pages} {220506} (\bibinfo {year} {2017})}\BibitemShut {NoStop}%
\bibitem [{\citenamefont {Peets}\ \emph {et~al.}(2019)\citenamefont {Peets},
  \citenamefont {Cheng}, \citenamefont {Ying}, \citenamefont {Kriener},
  \citenamefont {Shen}, \citenamefont {Li},\ and\ \citenamefont
  {Feng}}]{Peets2019}%
  \BibitemOpen
  \bibfield  {author} {\bibinfo {author} {\bibfnamefont {D.~C.}\ \bibnamefont
  {Peets}}, \bibinfo {author} {\bibfnamefont {E.}~\bibnamefont {Cheng}},
  \bibinfo {author} {\bibfnamefont {T.}~\bibnamefont {Ying}}, \bibinfo {author}
  {\bibfnamefont {M.}~\bibnamefont {Kriener}}, \bibinfo {author} {\bibfnamefont
  {X.}~\bibnamefont {Shen}}, \bibinfo {author} {\bibfnamefont {S.}~\bibnamefont
  {Li}}, \ and\ \bibinfo {author} {\bibfnamefont {D.}~\bibnamefont {Feng}},\
  }\href {\doibase 10.1103/PhysRevB.99.144519} {\bibfield  {journal} {\bibinfo
  {journal} {Phys. Rev. B}\ }\textbf {\bibinfo {volume} {99}},\ \bibinfo
  {pages} {144519} (\bibinfo {year} {2019})}\BibitemShut {NoStop}%
\bibitem [{\citenamefont {{Salis}}\ \emph {et~al.}(2018)\citenamefont
  {{Salis}}, \citenamefont {{Rodi{\`e}re}}, \citenamefont {{Leng}},
  \citenamefont {{Huang}},\ and\ \citenamefont {{de Visser}}}]{Salis2018}%
  \BibitemOpen
  \bibfield  {author} {\bibinfo {author} {\bibfnamefont {M.~V.}\ \bibnamefont
  {{Salis}}}, \bibinfo {author} {\bibfnamefont {P.}~\bibnamefont
  {{Rodi{\`e}re}}}, \bibinfo {author} {\bibfnamefont {H.}~\bibnamefont
  {{Leng}}}, \bibinfo {author} {\bibfnamefont {Y.~K.}\ \bibnamefont {{Huang}}},
  \ and\ \bibinfo {author} {\bibfnamefont {A.}~\bibnamefont {{de Visser}}},\
  }\href@noop {} {\bibfield  {journal} {\bibinfo  {journal} {J. Phys: Condens.
  Matter}\ }\textbf {\bibinfo {volume} {30}},\ \bibinfo {pages} {505602}
  (\bibinfo {year} {2018})}\BibitemShut {NoStop}%
\bibitem [{\citenamefont {Amit}\ and\ \citenamefont {Singh}(2018)}]{Amit2018}%
  \BibitemOpen
  \bibfield  {author} {\bibinfo {author} {\bibnamefont {Amit}}\ and\ \bibinfo
  {author} {\bibfnamefont {Y.}~\bibnamefont {Singh}},\ }\href {\doibase
  10.1103/PhysRevB.97.054515} {\bibfield  {journal} {\bibinfo  {journal} {Phys.
  Rev. B}\ }\textbf {\bibinfo {volume} {97}},\ \bibinfo {pages} {054515}
  (\bibinfo {year} {2018})}\BibitemShut {NoStop}%
\bibitem [{\citenamefont {Das}\ \emph {et~al.}(2018)\citenamefont {Das},
  \citenamefont {Amit}, \citenamefont {Sirohi}, \citenamefont {Yadav},
  \citenamefont {Gayen}, \citenamefont {Singh},\ and\ \citenamefont
  {Sheet}}]{Das2018}%
  \BibitemOpen
  \bibfield  {author} {\bibinfo {author} {\bibfnamefont {S.}~\bibnamefont
  {Das}}, \bibinfo {author} {\bibnamefont {Amit}}, \bibinfo {author}
  {\bibfnamefont {A.}~\bibnamefont {Sirohi}}, \bibinfo {author} {\bibfnamefont
  {L.}~\bibnamefont {Yadav}}, \bibinfo {author} {\bibfnamefont
  {S.}~\bibnamefont {Gayen}}, \bibinfo {author} {\bibfnamefont
  {Y.}~\bibnamefont {Singh}}, \ and\ \bibinfo {author} {\bibfnamefont
  {G.}~\bibnamefont {Sheet}},\ }\href {\doibase 10.1103/PhysRevB.97.014523}
  {\bibfield  {journal} {\bibinfo  {journal} {Phys. Rev. B}\ }\textbf {\bibinfo
  {volume} {97}},\ \bibinfo {pages} {014523} (\bibinfo {year}
  {2018})}\BibitemShut {NoStop}%
\bibitem [{\citenamefont {Sirohi}\ \emph {et~al.}(2019)\citenamefont {Sirohi},
  \citenamefont {Das}, \citenamefont {Adhikary}, \citenamefont {Chowdhury},
  \citenamefont {Vashist}, \citenamefont {Singh}, \citenamefont {Gayen},
  \citenamefont {Das},\ and\ \citenamefont {Sheet}}]{Sirohi2019}%
  \BibitemOpen
  \bibfield  {author} {\bibinfo {author} {\bibfnamefont {A.}~\bibnamefont
  {Sirohi}}, \bibinfo {author} {\bibfnamefont {S.}~\bibnamefont {Das}},
  \bibinfo {author} {\bibfnamefont {P.}~\bibnamefont {Adhikary}}, \bibinfo
  {author} {\bibfnamefont {R.~R.}\ \bibnamefont {Chowdhury}}, \bibinfo {author}
  {\bibfnamefont {A.}~\bibnamefont {Vashist}}, \bibinfo {author} {\bibfnamefont
  {Y.}~\bibnamefont {Singh}}, \bibinfo {author} {\bibfnamefont
  {S.}~\bibnamefont {Gayen}}, \bibinfo {author} {\bibfnamefont
  {T.}~\bibnamefont {Das}}, \ and\ \bibinfo {author} {\bibfnamefont
  {G.}~\bibnamefont {Sheet}},\ }\href {\doibase 10.1088/1361-648x/aaf49c}
  {\bibfield  {journal} {\bibinfo  {journal} {J. Phys: Condens. Matter}\
  }\textbf {\bibinfo {volume} {31}},\ \bibinfo {pages} {085701} (\bibinfo
  {year} {2019})}\BibitemShut {NoStop}%
\bibitem [{\citenamefont {{Teknowijoyo}}\ \emph {et~al.}(2018)\citenamefont
  {{Teknowijoyo}}, \citenamefont {{Jo}}, \citenamefont {{Scheurer}},
  \citenamefont {{Tanatar}}, \citenamefont {{Cho}}, \citenamefont {{Bud'ko}},
  \citenamefont {{Orth}}, \citenamefont {{Canfield}},\ and\ \citenamefont
  {{Prozorov}}}]{Teknowijoyo2018}%
  \BibitemOpen
  \bibfield  {author} {\bibinfo {author} {\bibfnamefont {S.}~\bibnamefont
  {{Teknowijoyo}}}, \bibinfo {author} {\bibfnamefont {N.~H.}\ \bibnamefont
  {{Jo}}}, \bibinfo {author} {\bibfnamefont {M.~S.}\ \bibnamefont
  {{Scheurer}}}, \bibinfo {author} {\bibfnamefont {M.~A.}\ \bibnamefont
  {{Tanatar}}}, \bibinfo {author} {\bibfnamefont {K.}~\bibnamefont {{Cho}}},
  \bibinfo {author} {\bibfnamefont {S.~L.}\ \bibnamefont {{Bud'ko}}}, \bibinfo
  {author} {\bibfnamefont {P.~P.}\ \bibnamefont {{Orth}}}, \bibinfo {author}
  {\bibfnamefont {P.~C.}\ \bibnamefont {{Canfield}}}, \ and\ \bibinfo {author}
  {\bibfnamefont {R.}~\bibnamefont {{Prozorov}}},\ }\href@noop {} {\bibfield
  {journal} {\bibinfo  {journal} {Phys. Rev. B}\ }\textbf {\bibinfo {volume}
  {98}},\ \bibinfo {pages} {024508} (\bibinfo {year} {2018})}\BibitemShut
  {NoStop}%
\bibitem [{\citenamefont {Saint-James}\ and\ \citenamefont
  {de~Gennes}(1963)}]{Saint-James&deGennes1963}%
  \BibitemOpen
  \bibfield  {author} {\bibinfo {author} {\bibfnamefont {D.}~\bibnamefont
  {Saint-James}}\ and\ \bibinfo {author} {\bibfnamefont {P.~G.}\ \bibnamefont
  {de~Gennes}},\ }\href@noop {} {\bibfield  {journal} {\bibinfo  {journal}
  {Phys. Lett.}\ }\textbf {\bibinfo {volume} {7}},\ \bibinfo {pages} {306}
  (\bibinfo {year} {1963})}\BibitemShut {NoStop}%
\bibitem [{\citenamefont {Kimura}\ \emph {et~al.}(2016)\citenamefont {Kimura},
  \citenamefont {Kabeya}, \citenamefont {Saitoh}, \citenamefont {Satoh},
  \citenamefont {Ogi}, \citenamefont {Ohsaki},\ and\ \citenamefont
  {Aoki}}]{Kimura2016}%
  \BibitemOpen
  \bibfield  {author} {\bibinfo {author} {\bibfnamefont {N.}~\bibnamefont
  {Kimura}}, \bibinfo {author} {\bibfnamefont {N.}~\bibnamefont {Kabeya}},
  \bibinfo {author} {\bibfnamefont {K.}~\bibnamefont {Saitoh}}, \bibinfo
  {author} {\bibfnamefont {K.}~\bibnamefont {Satoh}}, \bibinfo {author}
  {\bibfnamefont {H.}~\bibnamefont {Ogi}}, \bibinfo {author} {\bibfnamefont
  {K.}~\bibnamefont {Ohsaki}}, \ and\ \bibinfo {author} {\bibfnamefont
  {H.}~\bibnamefont {Aoki}},\ }\href {\doibase 10.7566/JPSJ.85.024715}
  {\bibfield  {journal} {\bibinfo  {journal} {J. Phys. Soc. Jpn}\ }\textbf
  {\bibinfo {volume} {85}},\ \bibinfo {pages} {024715} (\bibinfo {year}
  {2016})}\BibitemShut {NoStop}%
\bibitem [{\citenamefont {Wang}\ \emph {et~al.}(2005)\citenamefont {Wang},
  \citenamefont {Lortz}, \citenamefont {Paderno}, \citenamefont {Filippov},
  \citenamefont {Abe}, \citenamefont {Tutsch},\ and\ \citenamefont
  {Junod}}]{Wang2005}%
  \BibitemOpen
  \bibfield  {author} {\bibinfo {author} {\bibfnamefont {Y.}~\bibnamefont
  {Wang}}, \bibinfo {author} {\bibfnamefont {R.}~\bibnamefont {Lortz}},
  \bibinfo {author} {\bibfnamefont {Y.}~\bibnamefont {Paderno}}, \bibinfo
  {author} {\bibfnamefont {V.}~\bibnamefont {Filippov}}, \bibinfo {author}
  {\bibfnamefont {S.}~\bibnamefont {Abe}}, \bibinfo {author} {\bibfnamefont
  {U.}~\bibnamefont {Tutsch}}, \ and\ \bibinfo {author} {\bibfnamefont
  {A.}~\bibnamefont {Junod}},\ }\href {\doibase 10.1103/PhysRevB.72.024548}
  {\bibfield  {journal} {\bibinfo  {journal} {Phys. Rev. B}\ }\textbf {\bibinfo
  {volume} {72}},\ \bibinfo {pages} {024548} (\bibinfo {year}
  {2005})}\BibitemShut {NoStop}%
\bibitem [{\citenamefont {Auer}\ and\ \citenamefont
  {Ullmaier}(1973)}]{Auer&Ullmaier1973}%
  \BibitemOpen
  \bibfield  {author} {\bibinfo {author} {\bibfnamefont {J.}~\bibnamefont
  {Auer}}\ and\ \bibinfo {author} {\bibfnamefont {H.}~\bibnamefont
  {Ullmaier}},\ }\href {\doibase 10.1103/PhysRevB.7.136} {\bibfield  {journal}
  {\bibinfo  {journal} {Phys. Rev. B}\ }\textbf {\bibinfo {volume} {7}},\
  \bibinfo {pages} {136} (\bibinfo {year} {1973})}\BibitemShut {NoStop}%
\bibitem [{\citenamefont {Le}\ \emph {et~al.}(2019)\citenamefont {Le},
  \citenamefont {Yin}, \citenamefont {Feng}, \citenamefont {Huang},
  \citenamefont {Che}, \citenamefont {Li}, \citenamefont {Shi},\ and\
  \citenamefont {Lu}}]{Le2019}%
  \BibitemOpen
  \bibfield  {author} {\bibinfo {author} {\bibfnamefont {T.}~\bibnamefont
  {Le}}, \bibinfo {author} {\bibfnamefont {L.}~\bibnamefont {Yin}}, \bibinfo
  {author} {\bibfnamefont {Z.}~\bibnamefont {Feng}}, \bibinfo {author}
  {\bibfnamefont {Q.}~\bibnamefont {Huang}}, \bibinfo {author} {\bibfnamefont
  {L.}~\bibnamefont {Che}}, \bibinfo {author} {\bibfnamefont {J.}~\bibnamefont
  {Li}}, \bibinfo {author} {\bibfnamefont {Y.}~\bibnamefont {Shi}}, \ and\
  \bibinfo {author} {\bibfnamefont {X.}~\bibnamefont {Lu}},\ }\href@noop {}
  {\bibfield  {journal} {\bibinfo  {journal} {Phys. Rev. B}\ }\textbf {\bibinfo
  {volume} {99}},\ \bibinfo {pages} {180504} (\bibinfo {year}
  {2019})}\BibitemShut {NoStop}%
\bibitem [{\citenamefont {Amato}(1997)}]{Amato1997}%
  \BibitemOpen
  \bibfield  {author} {\bibinfo {author} {\bibfnamefont {A.}~\bibnamefont
  {Amato}},\ }\href {\doibase 10.1103/RevModPhys.69.1119} {\bibfield  {journal}
  {\bibinfo  {journal} {Rev. Mod. Phys.}\ }\textbf {\bibinfo {volume} {69}},\
  \bibinfo {pages} {1119} (\bibinfo {year} {1997})}\BibitemShut {NoStop}%
\bibitem [{\citenamefont {Yaounc}\ and\ \citenamefont {Dalmas~de
  R\'{e}otier}(2011)}]{Yaounc&Dalmas2011}%
  \BibitemOpen
  \bibfield  {author} {\bibinfo {author} {\bibfnamefont {A.}~\bibnamefont
  {Yaounc}}\ and\ \bibinfo {author} {\bibfnamefont {P.}~\bibnamefont {Dalmas~de
  R\'{e}otier}},\ }\href@noop {} {\emph {\bibinfo {title} {Muon spin rotation,
  relaxation and resonance; applications to condensed matter.}}}\ (\bibinfo
  {publisher} {Oxford University Press},\ \bibinfo {address} {Oxford},\
  \bibinfo {year} {2011})\BibitemShut {NoStop}%
\bibitem [{\citenamefont {Blundell}(1999)}]{Blundell1999}%
  \BibitemOpen
  \bibfield  {author} {\bibinfo {author} {\bibfnamefont {S.~J.}\ \bibnamefont
  {Blundell}},\ }\href {\doibase 10.1080/001075199181521} {\bibfield  {journal}
  {\bibinfo  {journal} {Contemp. Phys.}\ }\textbf {\bibinfo {volume} {40}},\
  \bibinfo {pages} {175} (\bibinfo {year} {1999})}\BibitemShut {NoStop}%
\bibitem [{\citenamefont {Gladisch}\ \emph {et~al.}(1979)\citenamefont
  {Gladisch}, \citenamefont {Herlach}, \citenamefont {Metz}, \citenamefont
  {Orth}, \citenamefont {zu~Putlitz}, \citenamefont {Seeger}, \citenamefont
  {Teichler}, \citenamefont {Wahl},\ and\ \citenamefont
  {Wigand}}]{Gladisch1979}%
  \BibitemOpen
  \bibfield  {author} {\bibinfo {author} {\bibfnamefont {M.}~\bibnamefont
  {Gladisch}}, \bibinfo {author} {\bibfnamefont {D.}~\bibnamefont {Herlach}},
  \bibinfo {author} {\bibfnamefont {H.}~\bibnamefont {Metz}}, \bibinfo {author}
  {\bibfnamefont {H.}~\bibnamefont {Orth}}, \bibinfo {author} {\bibfnamefont
  {G.}~\bibnamefont {zu~Putlitz}}, \bibinfo {author} {\bibfnamefont
  {A.}~\bibnamefont {Seeger}}, \bibinfo {author} {\bibfnamefont
  {H.}~\bibnamefont {Teichler}}, \bibinfo {author} {\bibfnamefont
  {W.}~\bibnamefont {Wahl}}, \ and\ \bibinfo {author} {\bibfnamefont
  {M.}~\bibnamefont {Wigand}},\ }\href {\doibase 10.1007/BF01028778} {\bibfield
   {journal} {\bibinfo  {journal} {Hyperfine Interact.}\ }\textbf {\bibinfo
  {volume} {6}},\ \bibinfo {pages} {109} (\bibinfo {year} {1979})}\BibitemShut
  {NoStop}%
\bibitem [{\citenamefont {Grebinnik}\ \emph {et~al.}(1980)\citenamefont
  {Grebinnik}, \citenamefont {Gurevich}, \citenamefont {Zhukov}, \citenamefont
  {Klimov}, \citenamefont {Levina}, \citenamefont {Maiorov}, \citenamefont
  {Manych}, \citenamefont {Mel'nikov}, \citenamefont {Nikol'skii},
  \citenamefont {Pirogov}, \citenamefont {Ponomarev}, \citenamefont {Roganov},
  \citenamefont {Selivanov},\ and\ \citenamefont {Suetin}}]{Grebinnik1980}%
  \BibitemOpen
  \bibfield  {author} {\bibinfo {author} {\bibfnamefont {V.}~\bibnamefont
  {Grebinnik}}, \bibinfo {author} {\bibfnamefont {I.}~\bibnamefont {Gurevich}},
  \bibinfo {author} {\bibfnamefont {V.}~\bibnamefont {Zhukov}}, \bibinfo
  {author} {\bibfnamefont {A.}~\bibnamefont {Klimov}}, \bibinfo {author}
  {\bibfnamefont {L.}~\bibnamefont {Levina}}, \bibinfo {author} {\bibfnamefont
  {V.}~\bibnamefont {Maiorov}}, \bibinfo {author} {\bibfnamefont
  {A.}~\bibnamefont {Manych}}, \bibinfo {author} {\bibfnamefont
  {E.}~\bibnamefont {Mel'nikov}}, \bibinfo {author} {\bibfnamefont
  {B.}~\bibnamefont {Nikol'skii}}, \bibinfo {author} {\bibfnamefont
  {A.}~\bibnamefont {Pirogov}}, \bibinfo {author} {\bibfnamefont
  {A.}~\bibnamefont {Ponomarev}}, \bibinfo {author} {\bibfnamefont
  {V.}~\bibnamefont {Roganov}}, \bibinfo {author} {\bibfnamefont
  {V.}~\bibnamefont {Selivanov}}, \ and\ \bibinfo {author} {\bibfnamefont
  {V.}~\bibnamefont {Suetin}},\ }\href@noop {} {\bibfield  {journal} {\bibinfo
  {journal} {Sov. Phys. JETP}\ }\textbf {\bibinfo {volume} {52}},\ \bibinfo
  {pages} {261} (\bibinfo {year} {1980})}\BibitemShut {NoStop}%
\bibitem [{\citenamefont {Egorov}\ \emph {et~al.}(2001)\citenamefont {Egorov},
  \citenamefont {Solt}, \citenamefont {Baines}, \citenamefont {Herlach},\ and\
  \citenamefont {Zimmermann}}]{Egerov2001}%
  \BibitemOpen
  \bibfield  {author} {\bibinfo {author} {\bibfnamefont {V.~S.}\ \bibnamefont
  {Egorov}}, \bibinfo {author} {\bibfnamefont {G.}~\bibnamefont {Solt}},
  \bibinfo {author} {\bibfnamefont {C.}~\bibnamefont {Baines}}, \bibinfo
  {author} {\bibfnamefont {D.}~\bibnamefont {Herlach}}, \ and\ \bibinfo
  {author} {\bibfnamefont {U.}~\bibnamefont {Zimmermann}},\ }\href {\doibase
  10.1103/PhysRevB.64.024524} {\bibfield  {journal} {\bibinfo  {journal} {Phys.
  Rev. B}\ }\textbf {\bibinfo {volume} {64}},\ \bibinfo {pages} {024524}
  (\bibinfo {year} {2001})}\BibitemShut {NoStop}%
\bibitem [{\citenamefont {{Aegerter}}\ \emph {et~al.}(2003)\citenamefont
  {{Aegerter}}, \citenamefont {{Keller}}, \citenamefont {{Lee}}, \citenamefont
  {{Ager}}, \citenamefont {{Ogrin}}, \citenamefont {{Cubitt}}, \citenamefont
  {{Forgan}}, \citenamefont {{Nutall}}, \citenamefont {{Kealey}}, \citenamefont
  {{Lloyd}}, \citenamefont {{Johnson}}, \citenamefont {{Riseman}},\ and\
  \citenamefont {{Nutley}}}]{Aegerter2003}%
  \BibitemOpen
  \bibfield  {author} {\bibinfo {author} {\bibfnamefont {C.~M.}\ \bibnamefont
  {{Aegerter}}}, \bibinfo {author} {\bibfnamefont {H.}~\bibnamefont
  {{Keller}}}, \bibinfo {author} {\bibfnamefont {S.~L.}\ \bibnamefont {{Lee}}},
  \bibinfo {author} {\bibfnamefont {C.}~\bibnamefont {{Ager}}}, \bibinfo
  {author} {\bibfnamefont {F.~Y.}\ \bibnamefont {{Ogrin}}}, \bibinfo {author}
  {\bibfnamefont {R.}~\bibnamefont {{Cubitt}}}, \bibinfo {author}
  {\bibfnamefont {E.~M.}\ \bibnamefont {{Forgan}}}, \bibinfo {author}
  {\bibfnamefont {W.~J.}\ \bibnamefont {{Nutall}}}, \bibinfo {author}
  {\bibfnamefont {P.~G.}\ \bibnamefont {{Kealey}}}, \bibinfo {author}
  {\bibfnamefont {S.~H.}\ \bibnamefont {{Lloyd}}}, \bibinfo {author}
  {\bibfnamefont {S.~T.}\ \bibnamefont {{Johnson}}}, \bibinfo {author}
  {\bibfnamefont {T.~M.}\ \bibnamefont {{Riseman}}}, \ and\ \bibinfo {author}
  {\bibfnamefont {M.~P.}\ \bibnamefont {{Nutley}}},\ }\href@noop {} {\bibfield
  {journal} {\bibinfo  {journal} {arXiv e-prints}\ } (\bibinfo {year}
  {2003})},\ \Eprint {http://arxiv.org/abs/cond-mat/0305595} {cond-mat/0305595
  [cond-mat.supr-con]} \BibitemShut {NoStop}%
\bibitem [{\citenamefont {{Kozhevnikov}}\ \emph {et~al.}(2018)\citenamefont
  {{Kozhevnikov}}, \citenamefont {{Suter}}, \citenamefont {{Prokscha}},\ and\
  \citenamefont {{Van Haesendonck}}}]{Kozhevnikov2018}%
  \BibitemOpen
  \bibfield  {author} {\bibinfo {author} {\bibfnamefont {V.}~\bibnamefont
  {{Kozhevnikov}}}, \bibinfo {author} {\bibfnamefont {A.}~\bibnamefont
  {{Suter}}}, \bibinfo {author} {\bibfnamefont {T.}~\bibnamefont {{Prokscha}}},
  \ and\ \bibinfo {author} {\bibfnamefont {C.}~\bibnamefont {{Van
  Haesendonck}}},\ }\href@noop {} {\bibfield  {journal} {\bibinfo  {journal}
  {arXiv e-prints}\ } (\bibinfo {year} {2018})},\ \Eprint
  {http://arxiv.org/abs/1802.08299} {1802.08299 [cond-mat.supr-con]}
  \BibitemShut {NoStop}%
\bibitem [{\citenamefont {Khasanov}\ \emph {et~al.}(2019)\citenamefont
  {Khasanov}, \citenamefont {Radonji\ifmmode~\acute{c}\else \'{c}\fi{}},
  \citenamefont {Luetkens}, \citenamefont {Morenzoni}, \citenamefont {Simutis},
  \citenamefont {Sch\"onecker}, \citenamefont {Appelt}, \citenamefont
  {\"Ostlin}, \citenamefont {Chioncel},\ and\ \citenamefont
  {Amato}}]{Khasanov2019}%
  \BibitemOpen
  \bibfield  {author} {\bibinfo {author} {\bibfnamefont {R.}~\bibnamefont
  {Khasanov}}, \bibinfo {author} {\bibfnamefont {M.~M.}\ \bibnamefont
  {Radonji\ifmmode~\acute{c}\else \'{c}\fi{}}}, \bibinfo {author}
  {\bibfnamefont {H.}~\bibnamefont {Luetkens}}, \bibinfo {author}
  {\bibfnamefont {E.}~\bibnamefont {Morenzoni}}, \bibinfo {author}
  {\bibfnamefont {G.}~\bibnamefont {Simutis}}, \bibinfo {author} {\bibfnamefont
  {S.}~\bibnamefont {Sch\"onecker}}, \bibinfo {author} {\bibfnamefont {W.~H.}\
  \bibnamefont {Appelt}}, \bibinfo {author} {\bibfnamefont {A.}~\bibnamefont
  {\"Ostlin}}, \bibinfo {author} {\bibfnamefont {L.}~\bibnamefont {Chioncel}},
  \ and\ \bibinfo {author} {\bibfnamefont {A.}~\bibnamefont {Amato}},\ }\href
  {\doibase 10.1103/PhysRevB.99.174506} {\bibfield  {journal} {\bibinfo
  {journal} {Phys. Rev. B}\ }\textbf {\bibinfo {volume} {99}},\ \bibinfo
  {pages} {174506} (\bibinfo {year} {2019})}\BibitemShut {NoStop}%
\bibitem [{\citenamefont {Karl}\ \emph {et~al.}(2019)\citenamefont {Karl},
  \citenamefont {Burri}, \citenamefont {Amato}, \citenamefont {Doneg\`a},
  \citenamefont {Gvasaliya}, \citenamefont {Luetkens}, \citenamefont
  {Morenzoni},\ and\ \citenamefont {Khasanov}}]{Karl2019}%
  \BibitemOpen
  \bibfield  {author} {\bibinfo {author} {\bibfnamefont {R.}~\bibnamefont
  {Karl}}, \bibinfo {author} {\bibfnamefont {F.}~\bibnamefont {Burri}},
  \bibinfo {author} {\bibfnamefont {A.}~\bibnamefont {Amato}}, \bibinfo
  {author} {\bibfnamefont {M.}~\bibnamefont {Doneg\`a}}, \bibinfo {author}
  {\bibfnamefont {S.}~\bibnamefont {Gvasaliya}}, \bibinfo {author}
  {\bibfnamefont {H.}~\bibnamefont {Luetkens}}, \bibinfo {author}
  {\bibfnamefont {E.}~\bibnamefont {Morenzoni}}, \ and\ \bibinfo {author}
  {\bibfnamefont {R.}~\bibnamefont {Khasanov}},\ }\href {\doibase
  10.1103/PhysRevB.99.184515} {\bibfield  {journal} {\bibinfo  {journal} {Phys.
  Rev. B}\ }\textbf {\bibinfo {volume} {99}},\ \bibinfo {pages} {184515}
  (\bibinfo {year} {2019})}\BibitemShut {NoStop}%
\bibitem [{\citenamefont {Drew}\ \emph {et~al.}(2006)\citenamefont {Drew},
  \citenamefont {Lee}, \citenamefont {Ogrin}, \citenamefont {Charalambous},
  \citenamefont {Bancroft}, \citenamefont {Paul}, \citenamefont {Takabatake},\
  and\ \citenamefont {Baines}}]{Drew2006}%
  \BibitemOpen
  \bibfield  {author} {\bibinfo {author} {\bibfnamefont {A.}~\bibnamefont
  {Drew}}, \bibinfo {author} {\bibfnamefont {S.}~\bibnamefont {Lee}}, \bibinfo
  {author} {\bibfnamefont {F.}~\bibnamefont {Ogrin}}, \bibinfo {author}
  {\bibfnamefont {D.}~\bibnamefont {Charalambous}}, \bibinfo {author}
  {\bibfnamefont {N.}~\bibnamefont {Bancroft}}, \bibinfo {author}
  {\bibfnamefont {D.~M.}\ \bibnamefont {Paul}}, \bibinfo {author}
  {\bibfnamefont {T.}~\bibnamefont {Takabatake}}, \ and\ \bibinfo {author}
  {\bibfnamefont {C.}~\bibnamefont {Baines}},\ }\href {\doibase
  https://doi.org/10.1016/j.physb.2005.11.072} {\bibfield  {journal} {\bibinfo
  {journal} {Physica B}\ }\textbf {\bibinfo {volume} {374-375}},\ \bibinfo
  {pages} {270 } (\bibinfo {year} {2006})}\BibitemShut {NoStop}%
\bibitem [{\citenamefont {Anand}\ \emph {et~al.}(2011)\citenamefont {Anand},
  \citenamefont {Hillier}, \citenamefont {Adroja}, \citenamefont {Strydom},
  \citenamefont {Michor}, \citenamefont {McEwen},\ and\ \citenamefont
  {Rainford}}]{Anand2011}%
  \BibitemOpen
  \bibfield  {author} {\bibinfo {author} {\bibfnamefont {V.~K.}\ \bibnamefont
  {Anand}}, \bibinfo {author} {\bibfnamefont {A.~D.}\ \bibnamefont {Hillier}},
  \bibinfo {author} {\bibfnamefont {D.~T.}\ \bibnamefont {Adroja}}, \bibinfo
  {author} {\bibfnamefont {A.~M.}\ \bibnamefont {Strydom}}, \bibinfo {author}
  {\bibfnamefont {H.}~\bibnamefont {Michor}}, \bibinfo {author} {\bibfnamefont
  {K.~A.}\ \bibnamefont {McEwen}}, \ and\ \bibinfo {author} {\bibfnamefont
  {B.~D.}\ \bibnamefont {Rainford}},\ }\href {\doibase
  10.1103/PhysRevB.83.064522} {\bibfield  {journal} {\bibinfo  {journal} {Phys.
  Rev. B}\ }\textbf {\bibinfo {volume} {83}},\ \bibinfo {pages} {064522}
  (\bibinfo {year} {2011})}\BibitemShut {NoStop}%
\bibitem [{\citenamefont {Anand}\ \emph {et~al.}(2014)\citenamefont {Anand},
  \citenamefont {Britz}, \citenamefont {Bhattacharyya}, \citenamefont {Adroja},
  \citenamefont {Hillier}, \citenamefont {Strydom}, \citenamefont {Kockelmann},
  \citenamefont {Rainford},\ and\ \citenamefont {McEwen}}]{Anand2014}%
  \BibitemOpen
  \bibfield  {author} {\bibinfo {author} {\bibfnamefont {V.~K.}\ \bibnamefont
  {Anand}}, \bibinfo {author} {\bibfnamefont {D.}~\bibnamefont {Britz}},
  \bibinfo {author} {\bibfnamefont {A.}~\bibnamefont {Bhattacharyya}}, \bibinfo
  {author} {\bibfnamefont {D.~T.}\ \bibnamefont {Adroja}}, \bibinfo {author}
  {\bibfnamefont {A.~D.}\ \bibnamefont {Hillier}}, \bibinfo {author}
  {\bibfnamefont {A.~M.}\ \bibnamefont {Strydom}}, \bibinfo {author}
  {\bibfnamefont {W.}~\bibnamefont {Kockelmann}}, \bibinfo {author}
  {\bibfnamefont {B.~D.}\ \bibnamefont {Rainford}}, \ and\ \bibinfo {author}
  {\bibfnamefont {K.~A.}\ \bibnamefont {McEwen}},\ }\href {\doibase
  10.1103/PhysRevB.90.014513} {\bibfield  {journal} {\bibinfo  {journal} {Phys.
  Rev. B}\ }\textbf {\bibinfo {volume} {90}},\ \bibinfo {pages} {014513}
  (\bibinfo {year} {2014})}\BibitemShut {NoStop}%
\bibitem [{\citenamefont {Smidman}\ \emph {et~al.}(2014)\citenamefont
  {Smidman}, \citenamefont {Hillier}, \citenamefont {Adroja}, \citenamefont
  {Lees}, \citenamefont {Anand}, \citenamefont {Singh}, \citenamefont {Smith},
  \citenamefont {Paul},\ and\ \citenamefont {Balakrishnan}}]{Smidman2014}%
  \BibitemOpen
  \bibfield  {author} {\bibinfo {author} {\bibfnamefont {M.}~\bibnamefont
  {Smidman}}, \bibinfo {author} {\bibfnamefont {A.~D.}\ \bibnamefont
  {Hillier}}, \bibinfo {author} {\bibfnamefont {D.~T.}\ \bibnamefont {Adroja}},
  \bibinfo {author} {\bibfnamefont {M.~R.}\ \bibnamefont {Lees}}, \bibinfo
  {author} {\bibfnamefont {V.~K.}\ \bibnamefont {Anand}}, \bibinfo {author}
  {\bibfnamefont {R.~P.}\ \bibnamefont {Singh}}, \bibinfo {author}
  {\bibfnamefont {R.~I.}\ \bibnamefont {Smith}}, \bibinfo {author}
  {\bibfnamefont {D.~M.}\ \bibnamefont {Paul}}, \ and\ \bibinfo {author}
  {\bibfnamefont {G.}~\bibnamefont {Balakrishnan}},\ }\href {\doibase
  10.1103/PhysRevB.89.094509} {\bibfield  {journal} {\bibinfo  {journal} {Phys.
  Rev. B}\ }\textbf {\bibinfo {volume} {89}},\ \bibinfo {pages} {094509}
  (\bibinfo {year} {2014})}\BibitemShut {NoStop}%
\bibitem [{\citenamefont {Singh}\ \emph {et~al.}(2019)\citenamefont {Singh},
  \citenamefont {Hillier},\ and\ \citenamefont {Singh}}]{Singh2019}%
  \BibitemOpen
  \bibfield  {author} {\bibinfo {author} {\bibfnamefont {D.}~\bibnamefont
  {Singh}}, \bibinfo {author} {\bibfnamefont {A.~D.}\ \bibnamefont {Hillier}},
  \ and\ \bibinfo {author} {\bibfnamefont {R.~P.}\ \bibnamefont {Singh}},\
  }\href {\doibase 10.1103/PhysRevB.99.134509} {\bibfield  {journal} {\bibinfo
  {journal} {Phys. Rev. B}\ }\textbf {\bibinfo {volume} {99}},\ \bibinfo
  {pages} {134509} (\bibinfo {year} {2019})}\BibitemShut {NoStop}%
\bibitem [{\citenamefont {Beare}\ \emph {et~al.}(2019)\citenamefont {Beare},
  \citenamefont {Nugent}, \citenamefont {Wilson}, \citenamefont {Cai},
  \citenamefont {Munsie}, \citenamefont {Amon}, \citenamefont {Leithe-Jasper},
  \citenamefont {Gong}, \citenamefont {Guo}, \citenamefont {Guguchia},
  \citenamefont {Grin}, \citenamefont {Uemura}, \citenamefont {Svanidze},\ and\
  \citenamefont {Luke}}]{Beare2019}%
  \BibitemOpen
  \bibfield  {author} {\bibinfo {author} {\bibfnamefont {J.}~\bibnamefont
  {Beare}}, \bibinfo {author} {\bibfnamefont {M.}~\bibnamefont {Nugent}},
  \bibinfo {author} {\bibfnamefont {M.~N.}\ \bibnamefont {Wilson}}, \bibinfo
  {author} {\bibfnamefont {Y.}~\bibnamefont {Cai}}, \bibinfo {author}
  {\bibfnamefont {T.~J.~S.}\ \bibnamefont {Munsie}}, \bibinfo {author}
  {\bibfnamefont {A.}~\bibnamefont {Amon}}, \bibinfo {author} {\bibfnamefont
  {A.}~\bibnamefont {Leithe-Jasper}}, \bibinfo {author} {\bibfnamefont
  {Z.}~\bibnamefont {Gong}}, \bibinfo {author} {\bibfnamefont {S.~L.}\
  \bibnamefont {Guo}}, \bibinfo {author} {\bibfnamefont {Z.}~\bibnamefont
  {Guguchia}}, \bibinfo {author} {\bibfnamefont {Y.}~\bibnamefont {Grin}},
  \bibinfo {author} {\bibfnamefont {Y.~J.}\ \bibnamefont {Uemura}}, \bibinfo
  {author} {\bibfnamefont {E.}~\bibnamefont {Svanidze}}, \ and\ \bibinfo
  {author} {\bibfnamefont {G.~M.}\ \bibnamefont {Luke}},\ }\href {\doibase
  10.1103/PhysRevB.99.134510} {\bibfield  {journal} {\bibinfo  {journal} {Phys.
  Rev. B}\ }\textbf {\bibinfo {volume} {99}},\ \bibinfo {pages} {134510}
  (\bibinfo {year} {2019})}\BibitemShut {NoStop}%
\bibitem [{\citenamefont {Lyons}\ \emph {et~al.}(1976)\citenamefont {Lyons},
  \citenamefont {Schleich},\ and\ \citenamefont {Wold}}]{Lyons1976}%
  \BibitemOpen
  \bibfield  {author} {\bibinfo {author} {\bibfnamefont {A.}~\bibnamefont
  {Lyons}}, \bibinfo {author} {\bibfnamefont {D.}~\bibnamefont {Schleich}}, \
  and\ \bibinfo {author} {\bibfnamefont {A.}~\bibnamefont {Wold}},\ }\href@noop
  {} {\bibfield  {journal} {\bibinfo  {journal} {Mat. Res. Bull.}\ }\textbf
  {\bibinfo {volume} {11}},\ \bibinfo {pages} {1155} (\bibinfo {year}
  {1976})}\BibitemShut {NoStop}%
\bibitem [{\citenamefont {Chen}\ \emph {et~al.}(1991)\citenamefont {Chen},
  \citenamefont {Brug},\ and\ \citenamefont {Goldfarb}}]{Chen1991}%
  \BibitemOpen
  \bibfield  {author} {\bibinfo {author} {\bibfnamefont {D.-X.}\ \bibnamefont
  {Chen}}, \bibinfo {author} {\bibfnamefont {J.~A.}\ \bibnamefont {Brug}}, \
  and\ \bibinfo {author} {\bibfnamefont {R.~B.}\ \bibnamefont {Goldfarb}},\
  }\href@noop {} {\bibfield  {journal} {\bibinfo  {journal} {IEEE Trans.
  Magn.}\ }\textbf {\bibinfo {volume} {27}},\ \bibinfo {pages} {3601} (\bibinfo
  {year} {1991})}\BibitemShut {NoStop}%
\bibitem [{\citenamefont {Pardo}\ \emph {et~al.}(2004)\citenamefont {Pardo},
  \citenamefont {Chen},\ and\ \citenamefont {Sanchez}}]{Pardo2004}%
  \BibitemOpen
  \bibfield  {author} {\bibinfo {author} {\bibfnamefont {E.}~\bibnamefont
  {Pardo}}, \bibinfo {author} {\bibfnamefont {D.-X.}\ \bibnamefont {Chen}}, \
  and\ \bibinfo {author} {\bibfnamefont {A.}~\bibnamefont {Sanchez}},\
  }\href@noop {} {\bibfield  {journal} {\bibinfo  {journal} {J. Appl. Phys.}\
  }\textbf {\bibinfo {volume} {96}},\ \bibinfo {pages} {5365} (\bibinfo {year}
  {2004})}\BibitemShut {NoStop}%
\bibitem [{\citenamefont {Pratt}(2000)}]{Pratt2000}%
  \BibitemOpen
  \bibfield  {author} {\bibinfo {author} {\bibfnamefont {F.~L.}\ \bibnamefont
  {Pratt}},\ }\href {\doibase http://dx.doi.org/10.1016/S0921-4526(00)00328-8}
  {\bibfield  {journal} {\bibinfo  {journal} {Physica B}\ }\textbf {\bibinfo
  {volume} {289}},\ \bibinfo {pages} {710 } (\bibinfo {year}
  {2000})}\BibitemShut {NoStop}%
\bibitem [{\citenamefont {Suter}\ and\ \citenamefont
  {Wojek}(2012)}]{Suter&Wojek2012}%
  \BibitemOpen
  \bibfield  {author} {\bibinfo {author} {\bibfnamefont {A.}~\bibnamefont
  {Suter}}\ and\ \bibinfo {author} {\bibfnamefont {B.}~\bibnamefont {Wojek}},\
  }\href {\doibase http://dx.doi.org/10.1016/j.phpro.2012.04.042} {\bibfield
  {journal} {\bibinfo  {journal} {Physics Procedia}\ }\textbf {\bibinfo
  {volume} {30}},\ \bibinfo {pages} {69 } (\bibinfo {year} {2012})}\BibitemShut
  {NoStop}%
\bibitem [{\citenamefont {Huebener}(1979)}]{Huebener1979}%
  \BibitemOpen
  \bibfield  {author} {\bibinfo {author} {\bibfnamefont {R.~P.}\ \bibnamefont
  {Huebener}},\ }\href@noop {} {\emph {\bibinfo {title} {Magnetic Flux
  Structures in Superconductors}}}\ (\bibinfo  {publisher} {Springer, Berlin},\
  \bibinfo {year} {1979})\BibitemShut {NoStop}%
\bibitem [{\citenamefont {Landau}(1937)}]{Landau1937}%
  \BibitemOpen
  \bibfield  {author} {\bibinfo {author} {\bibfnamefont {L.~D.}\ \bibnamefont
  {Landau}},\ }\href@noop {} {\bibfield  {journal} {\bibinfo  {journal} {Zh.
  Eksp. Teor. Fiz.}\ }\textbf {\bibinfo {volume} {7}},\ \bibinfo {pages} {371}
  (\bibinfo {year} {1937})}\BibitemShut {NoStop}%
\end{thebibliography}%
\bibliographystyle{apsrev4-1}

\end{document}